\documentclass[aps,twocolumn,superscriptaddress,showpacs,prb]{revtex4-1}

\usepackage{amsmath}
\usepackage{amssymb}
\usepackage{graphicx}
\usepackage[dvipsnames]{xcolor}
\usepackage{hyperref}
\usepackage{bm}

\begin{document}

\title{Nonequilibrium selection of magnetic order in driven triangular XY antiferromagnet}

\author{Yuan Wan}
\affiliation{Institute of Physics, Chinese Academy of Sciences, Beijing 100190, China}
\affiliation{Rudolf Peierls Centre for Theoretical Physics, University of Oxford, Oxford OX1 3NP, United Kingdom}
\author{Roderich Moessner}
\affiliation{Max-Planck Institute for the Physics of Complex Systems, D-00187 Dresden, Germany}

\begin{abstract}
We show that a weak periodic drive removes the accidental degeneracy in the ground state of the XY antiferromagnet on the triangular lattice in a uniform static magnetic field. The underlying mechanism involves adding a small periodically modulated component to the magnetic field, which influences finite-frequency modes to generate an effective potential  for the accidental pseudo-Goldstone mode. This selection can be arranged to compete with the degeneracy lifting via the thermal order by disorder mechanism. This yields a non-equilibrium phase transition as the relative strength of the two mechanisms is tuned by varying temperature. This proposal may be amenable to experimental realization, in particular as applying the field is noninvasive, and no complex bath engineering is required.
\end{abstract}

\date{\today}

%\pacs{75.10.-b,75.10.Hk,75.40.Mg}

\maketitle

\section{Introduction}

The driven-dissipative system is a versatile platform for studying non-equilibrium phenomena. In a driven-dissipative system, the energy flux from the drive to the thermal bath breaks detailed balance, thereby giving access to non-equilibrium phases and phase transitions. Since B\'{e}nard's early experimental investigation of the eponymous convection phenomenon~\cite{Benard:1901}, numerous non-equilibrium phases and phase transitions have been discovered in a wide range of driven-dissipative systems.

\emph{Driven frustrated magnets} provide an ideal setting to explore non-equilibrium phases and phase transitions. In a frustrated magnet, the conflicting exchange interactions often produce a continuous manifold of accidentally degenerate ground states. The slow drifting motion within the ground state manifold, corresponding to the pseudo-Goldstone modes, is governed by the fast motion of the normal modes that bring the system out of the manifold~\cite{Doucot:1998}. Such nonlinear coupling naturally provides a mechanism for dynamically stabilizing \emph{non-equilibrium magnetic orders}: The normal modes, when coherently driven by external stimuli such as magnetic field pulse or AC magnetic field, generate an effective potential in the ground state manifold, which dynamically lifts the degeneracy~\cite{Wan:2017}. In particular, if other competing mechanisms are absent, a vanishingly small driving field is sufficient to stabilize the ground state with minimal effective potential energy.

Coupling the driven system to a thermal bath establishes a non-equilibrium steady state at late time. The thermal fluctuations from the bath on their own tend to stabilize the ground states with minimal free energy through the order by thermal disorder mechanism~\cite{Villain:1980,Shender:1982,Kawamura:1984,Henley:1989,Moessner:1998}. The states that are selected by the driving and by the thermal fluctuations can be different. This can lead to a \emph{non-equilibrium phase transition} in the steady state. If the driving strength is below a certain threshold, the dynamical landscape due to driving is overwhelmed by the thermal free energy landscape, and hence the system exhibits the same magnetic order as in equilibrium. If the driving strength exceeds the threshold, the system exhibits non-equilibrium magnetic order stabilized dynamically by the periodic driving.

For the specific case of periodic driving, the aforementioned dynamical stabilization mechanism resembles Floquet engineering~\cite{Kapitsa:1951,Landau:1976,Grifoni:1998,Bukov:2015,Oka:2018}. From this perspective, the presence of frustration offers great tunability. At sufficiently low temperature $T$, the magnitude of variations of the free energy landscape scales with $T$. The driving threshold for stabilizing non-equilibrium magnetic orders therefore scales with $T$ and can in principle be made arbitrarily small by reducing $T$. This may be compared to a conventional magnet without frustration, for which the threshold in general scales with the typical interaction energy~\cite{Mitamura:2014,Takayoshi:2014,Sato:2016,Kitamura:2017,Claassen:2017}.

\begin{figure}
\includegraphics[width=\columnwidth]{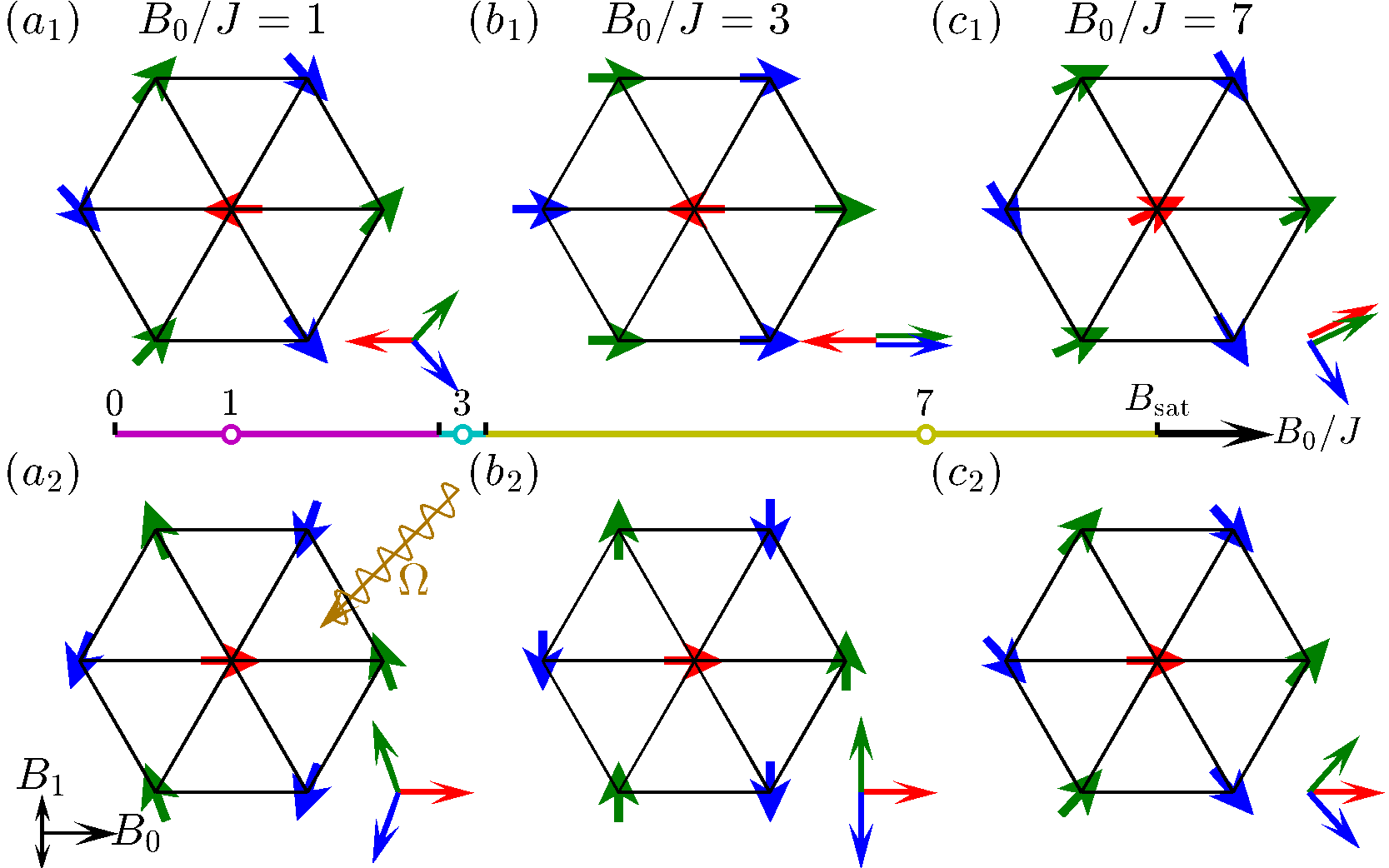}
\caption{Top row: Equilibrium magnetic orders in triangular XY antiferromagnet subject to a static, in-plane magnetic field ($B_0$) along the spin $x$ axis. XY spins in sublattice A, B, and C are colored in red, green, and blue.  Bottom row: Driving the system with a weak, time-periodic, in-plane magnetic field ($B_1$) orthogonal to $B_0$ dynamically stabilize the fan states. From left to right, the three columns show the specific spin configurations at $B_0/J=1,3,7$. Inset: Equilibrium phase diagram at fixed, low temperature $T$. As $B_0/J$ increases, the system is successively in the Y phase (magenta), up-up-down (UUD, cyan), 2:1 (yellow), and finally fully polarized phase (black), where all spins are aligned with the field.  Open circles mark the positions of the states given in (a$_1$),(b$_1$),(c$_1$) in the phase diagram. In a Y state (a$_1$), the spins in one sublattice are anti-aligned with the field, whereas the spins in the other two sublattices form symmetric angles with the field. In a UUD state (b$_1$), the spins in one sublattice are anti-aligned with the field, whereas the spins in the other two are aligned with the field. In a 2:1 state (c$_1$), the spins in one sublattice form one angle with the field, whereas the spins in the other two sublattices form a different angle with the field. In a fan state (a$_2$,b$_2$,c$_2$), spins in one of the three sublattices are aligned with the static field while spins in the other two sublattices form symmetric angles with the static field.}
\label{fig:sketch}
\end{figure}

In this work, we explore these ideas in a simple yet prominent frustrated spin model, namely the classical triangular XY antiferromagnet in a static, in-plane magnetic field~\cite{Kawamura:1984,Lee:1984,Lee:1986,Chubokov:1991}. Its classical Hamiltonian is given by,
\begin{align}
E = J\sum_{\langle ij\rangle}\cos(\phi_i-\phi_j)-B_0\sum_{i}\cos\phi_i,
\label{eq:model}
\end{align}
where $\phi_i$ is the polar angle of the XY spin on a triangular lattice site $i$. The first summation is over all nearest neighbor links, whereas the second is over all sites. The exchange constant $J>0$. The in-plane field $B_0$ is along the spin $x$ axis. Eq.~\eqref{eq:model} is the minimal model for easy-plane triangular antiferromagnets~\cite{Collins:1997} such as RbFe(MoO$_4$)$_2$~\cite{Inami:1996,Svistov:2003,Svistov:2006,Kenzelmann:2007,Smirnov:2007,Ribeiro:2011,Hearmon:2012,White:2013,Mitamura:2014}. It could also be realized by using Josephson junction arrays~\cite{Martinoli:2000} or cold atoms in an optical lattice~\cite{Struck:2013}.

The equilibrium phase diagram of Eq.~\eqref{eq:model} is well established~\cite{Lee:1984,Lee:1986}. When $B_0$ is below the saturation field $B_\mathrm{sat}=9J$, the ground states of  Eq.~\eqref{eq:model} are accidentally degenerate. At temperature $0<T\ll J$, thermal fluctuations lift the degeneracy through the order by thermal disorder mechanism. Fixing $T$ whilst increasing $B_0$ from 0 to $B_\mathrm{sat}$, the system is successively in the Y phase (Fig.~\ref{fig:sketch}(a$_1$)), the up-up-down (UUD) phase (Fig.~\ref{fig:sketch}(b$_1$)), and the 2:1 phase (Fig.~\ref{fig:sketch}(c$_1$)), all of which are stabilized by thermal fluctuations.

As we will show, one can select another kind of magnetic order, known as the fan states~\cite{Maryasin:2013,Smirnov:2017}, by driving the system with a time-periodic field $B_1$ along the spin $y$ axis (Fig.~\ref{fig:sketch}, bottom row). Although the fan states have the same symmetry as the Y states, the former are not related to the latter by symmetry. The fan states therefore constitute a distinct phase. The fan states are disfavored in thermal equilibrium as their free energy are the highest among the degenerate ground states. With periodic driving, the fan states become dynamically stable when $B_1$ exceeds a threshold value that is controlled by the temperature $T$. In what follows, we establish the non-equilibrium selection mechanism, its competition with thermal fluctuations, and the transition resulted from the competition thereof.

The rest of this work is organized as follows. In Sec.~\ref{sec:zeroT}, we analyze the dynamics of the model Eq.~\eqref{eq:model} at zero temperature and establish the selection of the fan states by periodic driving. In Sec.~\ref{sec:finiteT}, we show that, when coupled to a thermal bath, the competition between driving and thermal fluctuations gives rise to phase transitions between the fan states and the various equilibrium magnetic orders. In Sec.~\ref{sec:discuss}, we discuss the experimental feasibility of our proposal and a few open questions.

\section{\label{sec:zeroT} Dynamical selection at zero temperature}

\begin{figure*}
\includegraphics[width=\textwidth]{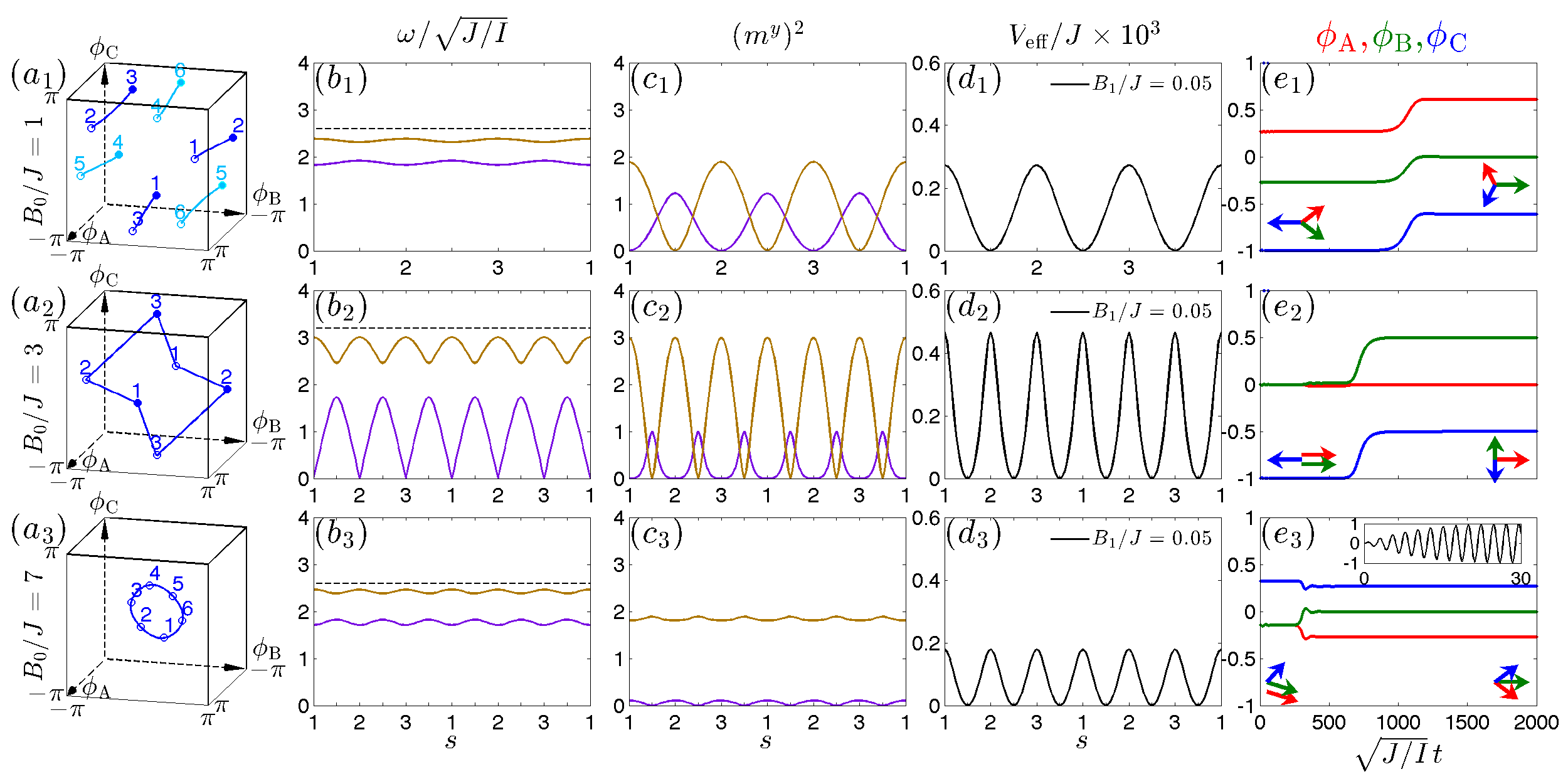}
\caption{Dynamical stabilization of the fan states due to periodic driving. The top, middle, and bottom rows respectively correspond to $B_0/J=1$, 3, and 7. Column (a): The one-dimensional degenerate ground state space (solid lines) in configuration space spanned by the sublattice spin angles $\phi_\mathrm{A,B,C}$. Columns (b) and (c): The intrinsic frequency $\omega$ of the two optical magnons and their magnetic dipole moments in the $y$ direction $m^y$ as functions of the ground state coordinate $s$. High and low frequency modes are colored in gold and purple, respectively. Dashed lines mark the driving frequency. Column (d): Driving-induced effective potential $V_\mathrm{eff}$. Driving amplitude $B_1=0.05J$ for all three cases. Column (e): Time-evolution of the sublattice spin orientation $\phi_\mathrm{A,B,C}$ after the driving is switched on at time $t=0$. Colored arrows show the sketch of the magnetic orders. Top row: the degenerate ground state space at $B_0/J=1$ contains two connected components (dark and light blue). $1\sim6$ mark the Y states. Circles with the same label mark identical states thanks to the periodicity $\phi\to\phi+2\pi$. Filled (open) circles are on the front (back) surfaces of the box. Fan states are at midpoints of two neighboring Y states. Results in (b$\sim$d) are for the component colored in darker blue. Driving frequency $\Omega = 2.6\sqrt{J/I}$. Middle row: The ground state space at $B_0/J=3$ contains topological singular points at the UUD states, labeled as $1\sim3$, where the ground state space self-intersects (a$_2$). Fan states are midpoints of two neighboring UUD states. The driving frequency $\Omega = 3.2\sqrt{IJ}$. Bottom row: the ground state space at $B_0/J=7$ has a single connected component. $1\sim6$ label the 2:1 states. The fan states are the midpoints of two neighboring 2:1 states. The driving frequency $\Omega = 2.6\sqrt{J/I}$. The time-profile of the periodic driving field is shown in the inset of (e$_3$).}
\label{fig:zeroT}
\end{figure*}

In this section, we analyze the dynamics of  Eq.~\eqref{eq:model} at $T=0$. We endow the XY spins with rotor dynamics. It is convenient to partition the triangular lattice into three sublattices, dubbed A, B, and C (Fig.~\ref{fig:sketch}).  $B_1$ couples to the uniform mode in each sublattice linearly and all other modes nonlinearly. To the leading order of the driving strength $B_1/J$, we may retain only the uniform modes and discard the rest. The validity of this approximation will also be justified \textit{a posteriori} by a comparison with direct numerical simulations. Within our approximation, the spins in the same sublattice take the same orientation, which we parametrize by polar angles $\phi_\mathrm{A,B,C}$. This reduces the many-body dynamics of Eq.~\eqref{eq:model} to a dynamical system with 3 degrees of freedom. Its Lagrangian is given by:
\begin{subequations}\label{eq:lagrangian}
\begin{align}
L &= \frac{NI}{6}\sum_{\alpha}\dot{\phi}^2_\alpha-\frac{NJ}{2}\sum_{\alpha\neq\beta}\cos(\phi_\alpha-\phi_\beta) \nonumber\\
& +\frac{N}{3}\sum_{\alpha}[B_0\cos\phi_\alpha+B_1(t)\sin\phi_\alpha].
\end{align}
Here, $N$ is the total number of spins. $I>0$ is the rotational inertia. $\alpha,\beta$ run over sublattice labels. $B_1(t)$ is a time-harmonic field: $B_1(t) = B_1\cos(\Omega t)$. We include damping through the Rayleigh dissipation function~\cite{Goldstein:1980}:
\begin{align}
R = \frac{Nk}{6}\sum_\alpha \dot{\phi}^2_\alpha,
\end{align}
\end{subequations}
where $k>0$ is the damping constant. The equations of motion are then obtained by using the Euler-Lagrange equation $d/dt(\partial L/\partial\dot{\phi}_\alpha)-\partial L/\partial\phi_\alpha = \partial R/\partial\dot{\phi}_\alpha$. Eq.~\eqref{eq:lagrangian} completes the description of the dynamical system, which will be the focus of the remaining part of this section.

In a ground state, $\phi_{A,B,C}$ satisfy $\sum_\alpha\cos\phi_\alpha=B_0/(3J)$ and $\sum_\alpha\sin\phi_\alpha=0$, where $\alpha$ runs over sublattices. These two conditions define a 1 dimensional ground state space embedded in 3 dimensional configuration space spanned by $\phi_\mathrm{A,B,C}$. The specific geometry and topology of the ground state space depends on $B_0/J$. Neglecting for the moment potential subtleties associated with topology, we may view the ground state space as a curve. It is therefore natural to parametrize the ground states by using the arc-length coordinate $s$~\cite{DoCarmo:2016}. To this end, we designate some ground state as the reference state. Each ground state is then parametrized by the length of the arc, $s$, that connects it to the said reference state. In particular, $s=0$ for the reference state. Details of the arc length parametrization are given in Appendix~\ref{app:arclength}.

The tangent vector of the ground state space, or the pseudo-Goldstone mode, corresponds to the motion within the ground state space. Orthogonal to the pseudo-Goldstone modes are the two normal modes, which correspond to deviations away from the ground state space. The driving field $B_1(t)$ forces the normal modes to oscillate with frequency $\Omega$. Meanwhile, the system may also drift slowly within the ground state space. We therefore postulate the following \emph{variational ansatz}:
\begin{align}
\label{eq:ansatz}
\phi_\alpha(t) = \phi^{(0)}_\alpha[s(t)]+\mathrm{Re}[A_\alpha(t) e^{i\Omega t}].
\end{align}
Here, $\phi^{(0)}_\alpha$ is the spin angle in the ground state. It depends on time through $s(t)$. $A_\alpha$ is the complex oscillation amplitude that evolves slowly in time. Based on the aforementioned picture, $\dot{s}/s,\dot{A}_\alpha/A_\alpha \ll \Omega$. We are interested in the weak driving regime $B_1/J\ll1$, which implies that $A_\alpha\ll1$.

We find the explicit time dependence of $s$ and $A_\alpha$ by using the method of averaged Lagrangian~\cite{Whitham:1974}. We substitute Eq.~\eqref{eq:ansatz} into Eq.~\eqref{eq:lagrangian}, expand the Lagrangian to the leading order in $A_\alpha$, and average the Lagrangian over a time-period of $2\pi/\Omega$. This procedure yields an averaged Lagrangian $\overline{L}$ that describes the slow dynamics of $s$ and $A_\alpha$. The equation of motion for $s$ is then obtained from $\overline{L}$. After some lengthy calculations (see Appendix \ref{app:derivation} for details), we obtain:
\begin{align}
\frac{NI}{3}\ddot{s}+\frac{Nk}{3}\dot{s} = -N\frac{\partial V_\mathrm{eff}}{\partial s}.
\end{align}
The effective potential $V_\mathrm{eff}$ admits a simple, analytical expression when the damping coefficient $k\ll \sqrt{IJ}$:
\begin{align}
V_\mathrm{eff}(s) = \frac{B^2_1}{12I}\sum_\lambda\frac{(\Omega^2-\omega^2_\lambda)}{(\omega^2_\lambda-\Omega^2)^2+k^2\Omega^2/I^2}(m^y_\lambda)^2,
\label{eq:Veff}
\end{align}
where the summation is over the two normal modes labeled by $\lambda$. $\omega_\lambda$ and $m^y_\lambda$ are respectively the intrinsic frequency and $y$-direction magnetic dipole moment of the mode $\lambda$. $V_\mathrm{eff}$ receives its $s$ dependence through $\omega_\lambda$ and $m^y_\lambda$. 

Eq.~\eqref{eq:Veff} is the central result of this section. It shows that the periodic driving induces an effective potential in the ground state space~\cite{Wan:2017}. It is important to bear in mind that we have made a series of assumptions in deriving Eq.~\eqref{eq:Veff} from Eq.~\eqref{eq:lagrangian}: (a) The driving amplitude is weak, $B_1\ll J$; (b) The driving frequency $\Omega$ is not in resonance with any of the two normal modes; (c) The normal mode frequencies are bounded from below above zero, $\omega_\lambda>0$; (d) The ground state space does not contain topological singularities; (e) The damping coefficient is relatively small, $k/\sqrt{IJ}\ll 1$. We also note that our approach differs from the usual mathematical framework of Floquet engineering in that the latter relies on the Magnus expansion~\cite{Grifoni:1998,Bukov:2015,Oka:2018}.

In what follows, we apply Eq.~\eqref{eq:Veff} to various representative cases and examine the selection effect. Specifically, we consider three representative values of $B_0/J$: $B_0/J=1$, 3, and 7. At thermal equilibrium, these parameters respectively put the system in the Y phase, the UUD phase, and the 2:1 phase (Fig.~\ref{fig:sketch}, inset). Here, we show that the periodic driving dynamically stabilizes the fan phase in all three cases.

We first focus on the case with $B_0/J=1$. In this case, the ground state space consists of two connected components (Fig.~\ref{fig:zeroT}(a$_1$)). The ground states in each component are connected by continuous rotation of spins. The two components are related to each other by mirror reflection with respect to spin-$x$ axis. Moving from one component to the other must overcome a high energy barrier. We henceforth neglect such a process and focus on only one connected component.

Fig.~\ref{fig:zeroT}(b$_1$) and (c$_1$) show the dependence of the intrinsic frequency $\omega_\lambda$ and the magnetic dipole moment in $y$ direction $m^y_\lambda$ on the ground state coordinate $s$. Fig.~\ref{fig:zeroT}(d$_1$) shows the effective potential $V_\mathrm{eff}(s)$ for the driving amplitude $B_1/J = 0.05$ and frequency $\Omega/\sqrt{J/I} = 2.6$. Throughout this work, we set the damping coefficient $k/\sqrt{IJ}=0.05$, a typical value for spin systems~\cite{Eriksson:2017}. The functional form of $V_\mathrm{eff}(s)$ essentially follows $(m^y)^2$ of the high frequency mode as it is close to resonance with $\Omega$. Crucially, the minima of $V_\mathrm{eff}$ are located at the fan states, which are midpoints of two neighboring Y states. Thus, periodic driving selects the fan states instead of the Y states.

We understand the dynamical stabilization of the fan states through the following heuristic picture. The high frequency normal mode is close to resonance with the drive and thus dominates the system's dynamical response. The fan states minimize the magnetic dipole moment of the high frequency mode along the direction of the driving field. This would minimize the system's \emph{dynamical} response, and consequently minimize the total kinetic energy. Our picture is reminiscent of Henley's argument for the selection of collinear magnetic order in $J_1$-$J_2$ square antiferromagnet~\cite{Henley:1989}. We stress that the dynamical stabilization is the opposite of the usual field selection effect, where the system prefers to maximize the \emph{static} response to the external field.

Note the effective potential $V_\mathrm{eff}$ depends on the driving frequency $\Omega$. Here, we have set $\Omega$ to be slightly above the high frequency mode, i.e. the band top of the spin wave spectrum. Our choice is based on two considerations. First, if one instead set $\Omega$ to be slightly below the band top, the minima of $V_\mathrm{eff}$ are located at the Y states rather than fan states. Second, while setting $\Omega$ to be in resonance with the high frequency mode may seem to enhance the magnitude of $V_\mathrm{eff}$, undesired nonlinear couplings, omitted in the present analysis, in fact destroy the dynamical stabilization effect.

Returning to the present choice of driving frequency $\Omega/\sqrt{IJ} = 2.6$, the selection of fan states is confirmed by a direct numerical simulation, where we integrate the many-body equation of motion for a system of $42\times42$ spins with periodic boundary conditions. We set the initial spin configuration to be a Y state. Since Y states are $V_\mathrm{eff}$ maxima (Fig.~\ref{fig:zeroT}(d$_1$)), we trigger their instability by assigning to each spin a random initial velocity $|\dot{\phi}_i|<10^{-10}\sqrt{J/I}$. After the driving field is ramped up over about 10 cycles of oscillation (Fig.~\ref{fig:zeroT}(e$_3$), inset), the system settles in a fan state by collective spin rotation (Fig.~\ref{fig:zeroT}(e$_1$)). Recall that the ground state space consists of two connected components. Starting from a specific Y state, only the three fan states belonging to the same connected component are accessible this way. 

Having demonstrated the selection of fan states over the Y states, we now turn to other regimes of $B_0/J$. Similar analyses shows that periodic driving stabilizes the fan states for the cases $B_0/J=3$ and 7 despite distinct topological properties of the ground state space (Fig.~\ref{fig:zeroT}, middle and bottom rows). Note, for $B_0/J=3$, the ground state space self-intersects at the UUD states (Fig.~\ref{fig:zeroT}(a$_2$)). In other words, the UUD states are the singular points of the ground state space. This is also manifest in the intrinsic frequencies of the normal modes (Fig.~\ref{fig:zeroT}(b$_2$)), where the low frequency mode softens as the system approaches the UUD states. Thus, Eq.~\eqref{eq:Veff} is inapplicable in the vicinity of the UUD states as the conditions (c) and (d) are violated. Nonetheless, the stability of the fan states inferred from Eq.~\eqref{eq:Veff} is robust. This is confirmed by direct numerical simulation, which shows the periodic drive dynamically stabilizes the fan states over the UUD states (Fig.~\ref{fig:zeroT}(e$_2$)).

\section{\label{sec:finiteT} Phase transitions at finite temperature}

\begin{figure*}
\includegraphics[width=\textwidth]{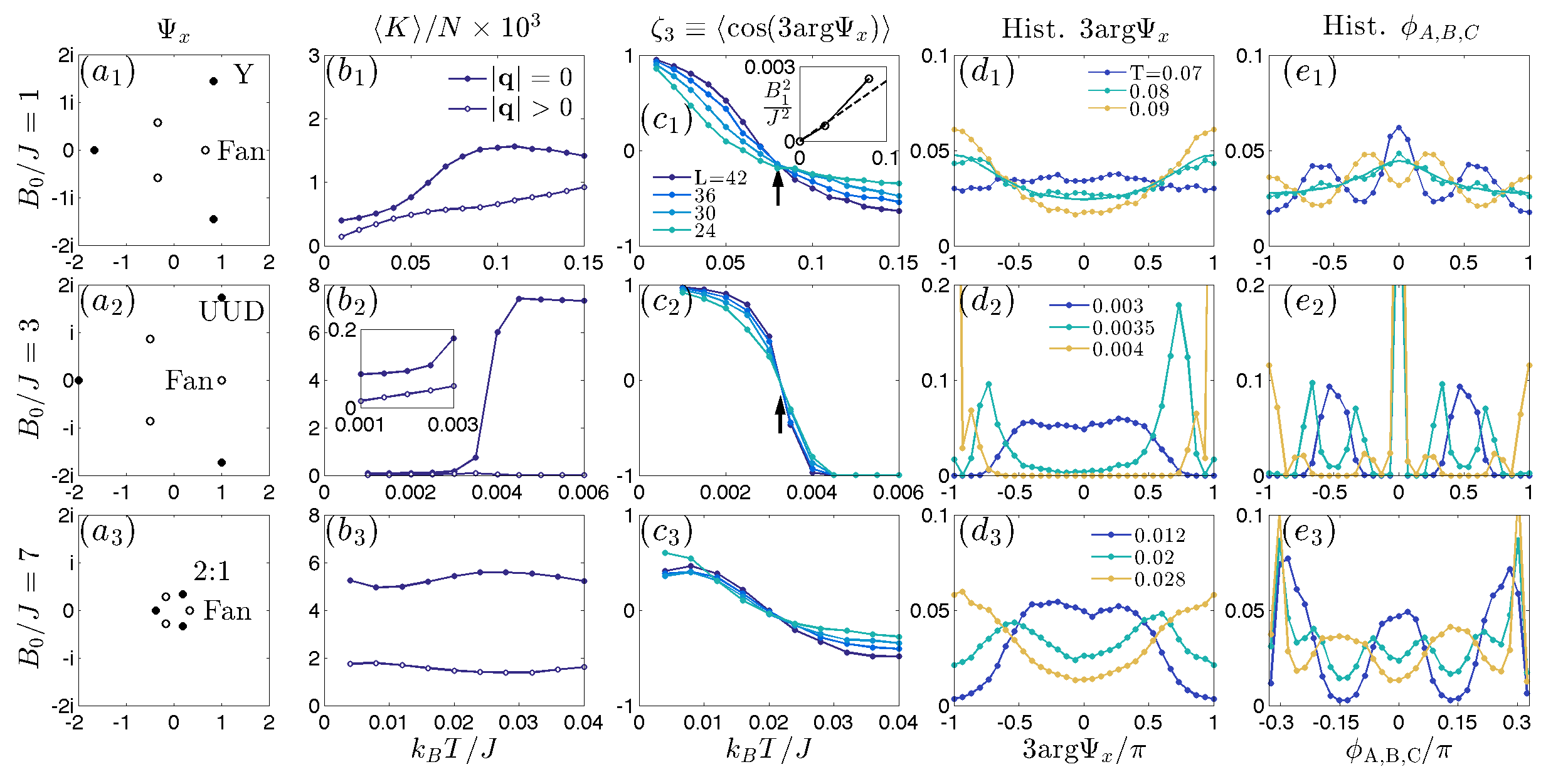}
\caption{Non-equilibrium phase transitions resulted from the competition between periodic drive and thermal fluctuations. Top, middle, and bottom rows show the results with static fields $B_0/J=1,3,7$, respectively. We set the driving amplitude $B_1=0.05J$ unless stated otherwise. Column (a): Value of order parameter $\Psi_x$ on the complex plane for the fan state (open circles) and the competing equilibrium orders (closed circles). Column (b):  Average kinetic energy density $\langle K\rangle/N$ as a function of bath temperature $T$. The kinetic energy density due to the wave vector $|\mathbf{q}|=0$ modes (closed circles) and $|\mathbf{q}|>0$ modes (open circles) are shown separately. For $|\mathbf{q}|>0$ modes, the thermal contribution $T/2$ is subtracted to highlight the heating effect. Column (c): Order parameter $\zeta_3$ as a function of $T$ for different system sizes $L$. Columns (d) and (e): Histograms of the order parameter angle $3\mathrm{arg}\Psi_x$ and the sublattice magnetization orientation $\phi_\mathrm{A,B,C}$ near the transition. The same color code corresponds to the same temperature. Note we do not distinguish the three sublattices in the latter histogram. Top row: Driving frequency $\Omega = 2.6\sqrt{IJ}$. The arrow in (c$_1$) marks the crossing point of the order parameter curves for different system sizes. The inset of (c$_1$) shows the scaling of $T_c$ with $B^2_1$ (open circles). The dashed line is the linear scaling relationship based on the crude estimate given in the main text. In (d$_1$) and (e$_1$), the solid cyan line shows the histogram of $3\mathrm{arg}\Psi_x$ and $\phi_\mathrm{A,B,C}$ obtained by randomly drawing ground states with uniform probability density. Middle row: $\Omega = 3.2\sqrt{IJ}$. The inset of (b$_2$) shows the enlarged view of the low-temperature part of the data. Arrow in (c$_2$) marks the crossing point of data curves for different system sizes. Bottom row: Driving frequency $\Omega = 2.6\sqrt{IJ}$.}
\label{fig:finiteT}
\end{figure*}

At finite temperature $T$, the effective potential $V_\mathrm{eff}$ must compete with the thermal fluctuations as the later disfavor the fan states. In this section, we investigate such competition numerically. To this end, we couple our system to a thermal bath at temperature $T$. The bath takes away the energy deposited by the periodic drive and establishes a non-equilibrium steady state at late time. 

Microscopically, the bath is modeled as Gaussian stochastic torque $\xi_i$, characterized by the correlation function $\langle\xi_i(t)\xi_j(t')\rangle = \sqrt{2kIk_BT}\delta_{ij}\delta(t-t')$. This leads to the standard Langevin equation~\cite{Loft:1987}:
\begin{subequations}\label{eq:langevin}
\begin{align}
I\ddot{\phi}_i+k\dot{\phi}_i = \tau_i+\xi_i,
\end{align}
where the mechanical torque $\tau_i$ is given by:
\begin{align}
\tau_i = J\sum_{j\in N_i}\sin(\phi_i-\phi_j)-B_0\sin\phi_i+B_1(t)\cos\phi_i.
\end{align}
\end{subequations}
Here, the summation is over all nearest neighbors of the site $i$. In particular, setting $T=0$ reduces Eq.~\eqref{eq:langevin} to the zero temperature many-body dynamics problem we have analyzed in Sec.~\ref{sec:zeroT}.  

We integrate Eq.~\eqref{eq:langevin} for a system of $L\times L$ spins with periodic boundary conditions by using the Bussi-Parrinello algorithm~\cite{Bussi:2007}. The details of the algorithm are given in Appendix~\ref{app:algorithm}. Results are presented for $L=42$ unless stated otherwise. Animations of the typical simulation runs can be found in the supplemental material, which shows how the system may be driven from the equilibrium phases to the non-equilibrium phase at early time~\cite{*[{See Supplemental Material at }]  [{ for the animations that show the evolution of the spin configuration in real time.}] SupplMat}. At late time, the system enters a synchronized state with discrete time-translation symmetry, $t\to t+2\pi/\Omega$. We record the system configurations at stroboscopic times $t_i$ when the driving field $B_1(t_i)=0$. Due to the discrete time-translation symmetry, this is effectively sampling from the same ensemble. At these instants, the symmetries of the driven system's Hamiltonian are the same as the equilibrium.  All averages are then performed within the stroboscopic ensemble.

To begin, we consider $B_0/J=1$. Similar to $T=0$, we set $B_1/J=0.05$ and $\Omega/\sqrt{J/I} = 2.6$. As $T$ increases from 0, We expect a transition from the fan phase to the Y phase. We therefore construct an order parameter to distinguish these two phases. Consider a complex order parameter $\Psi_x$ defined as $\mathrm{Re}\Psi_x \equiv m^x_\mathrm{A}-(m^x_\mathrm{B}+m^x_\mathrm{C})/2$ and $\mathrm{Im}\Psi_x \equiv \sqrt{3}/2(m^x_B-m^x_C)$, where $m^x_\mathrm{A,B,C}$ are the $x$-component of the magnetization density for sublattice A, B, and C, respectively~\cite{Lee:1984,Lee:1986}. In the absence of fluctuations, the complex argument angle of $\Psi_x$ takes value $0,\pm2\pi/3$ if the system is in fan phase and $\pi,\pm\pi/3$ if in Y phase (Fig.~\ref{fig:finiteT}(a$_1$)). Thus, $\zeta_3 \equiv \langle\cos(3\mathrm{arg}\Psi_x)\rangle$ is an order parameter that distinguishes the two phases: $\zeta_3 \to 1(-1)$ for the fan (Y) phase.

We first study the heating effect, which is inevitable due to the nonlinear coupling between the spin wave modes. We distinguish the modes with wave vector $\mathbf{q}=0$ and $\mathbf{q}\neq 0$. The former are coherently driven by $B_1$ and thus possesses a finite kinetic energy \emph{density} on their own ($\sim10^{-3}J$), which is one order of magnitude less than the equilibrium kinetic energy density $k_BT/2$ (Fig.~\ref{fig:finiteT}(b$_1$), closed circles). By contrast, the latter are dominated by the thermal fluctuations. Due to driving, their kinetic energy density is slightly larger than the thermal equilibrium value $k_BT/2$ (Fig.~\ref{fig:finiteT}(b$_1$), open circles). The small excess ($<10^{-3}J$) reflects moderate heating.

We then turn to the competition between the periodic driving and the thermal fluctuations. As $T/J$ increases from $0.01$ to $0.15$, the order parameter $\zeta_3$ passes from a positive value to a negative value, indicating a transition from a dynamically stabilized fan phase to a thermally stabilized Y phase (Fig.~\ref{fig:finiteT}(c$_1$)). The $\langle\zeta_3\rangle$ curves for various system sizes $L$ cross at approximately the same temperature $T_c$, which we interpret as the transition temperature. Using data for $L=36$ and 42, we estimate $T_c\approx 8.04\times10^{-2}J$.

$T_c$ is controlled by the driving amplitude. On one hand, for fixed driving frequency $\Omega$, Eq.~\eqref{eq:Veff} shows the magnitude of $V_\mathrm{eff}$ scales with $B^2_1/J$. On the other hand, the magnitude of the thermal free energy landscape is $T\Delta S$, where $\Delta S$ is the entropy difference between the fan states and Y states. At $T_c$, we expect the effective potential $V_\mathrm{eff}$ and the free energy landscape are comparable in magnitude, $B^2_1/J\sim  T_c\Delta S$, which yields $T_c\sim B^2_1$. Simulation indeed shows that $T_c$ roughly scales linearly with $B^2_1/J$ (Fig.~\ref{fig:finiteT}(c$_1$), inset).

To clarify the nature of the transition, we plot the histogram of the quantity $3\mathrm{arg}\Psi_x$  near $T_c$ (Fig.~\ref{fig:finiteT}(d$_1$)).  At $T/J=0.08$, which is in the vincinity of $T_c$, the histogram is well approximated by a toy model in which we randomly draw ground states from the degenerate ground state space with uniform probability. Note the resulting $3\mathrm{arg}\Psi_x$ histogram is not flat because $\mathrm{arg}\Psi_x$ is not uniform in the ground state space. The agreement between the simulation data and our toy model suggests the driving-induced $V_\mathrm{eff}$ and the thermal free energy landscape approximately cancel, leading to an almost uniform distribution of ground states. Below or above $T_c$, the histogram of $3\mathrm{arg}\Psi_x$ develops peaks at $0$ or $\pi$, corresponding to respectively the fan state and Y state.

The histogram of the polar angle of the sublattice magnetization $\phi_\mathrm{A,B,C}$  (Fig.~\ref{fig:finiteT}(e$_1$)) mirrors the behavior of the $3\mathrm{arg}\Psi_x$ histogram. At $T/J=0.08$, the histogram can be approximated by the aforementioned toy model. Above $T_c$, the histogram develops three peaks that correspond to the three spin angles in the fan phase (Fig.~\ref{fig:sketch}(a$_2$)). Likewise, below $T_c$, the histogram develops three peaks that correspond to the Y phase (Fig.~\ref{fig:sketch}(a$_1$)).

\begin{figure}
\includegraphics[width=\columnwidth]{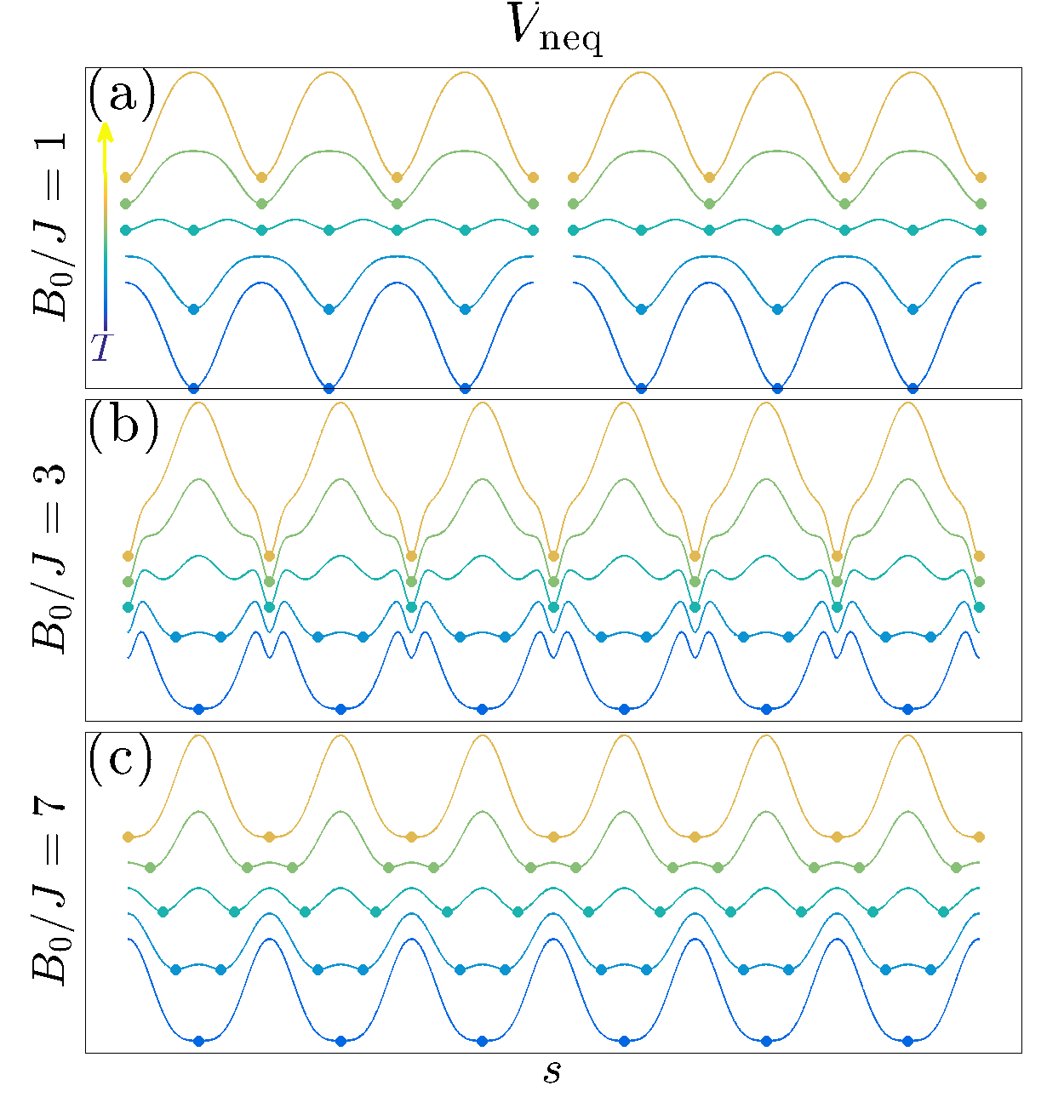}
\caption{Sketch of the evolution of the non-equilibrium effective energy landscape, $V_\mathrm{neq}$, as a function of the ground state coordinate $s$. The curves are shifted vertically or better visibility. Filled circles mark the minima of $V_\mathrm{neq}$. (a) Evolution of $V_\mathrm{neq}$ for $B_0/J=1$. Both connected components of the ground state space are shown. The minima jump from fan states to Y states as $T$ increases across $T_c$. At $T_c$, the fan states and Y states are degenerate, and the overall magnitude of $V_\mathrm{neq}$ is also relatively small. (b) Same as (a) but for $B_0/J=3$. As $T$ increases, the minima of $V_\mathrm{neq}$, originally at fan states at low $T$, split into pairs. Meanwhile, $V_\mathrm{eff}$ develop metastable minima at the UUD states, which eventually become the global minima. (c) Same as (a) but for $B_0/J=7$. The splitting of minima is similar to (b), but $V_\mathrm{neq}$ shows no metastable states.}
\label{fig:vneq}
\end{figure}

The evolution of the histograms may be heuristically understood as follows (Fig~\ref{fig:vneq}(a)). The competition between the periodic driving and the thermal fluctuations give rise to a non-equilibrium effective energy landscape in the ground state space, $V_\mathrm{neq}$, which interpolates the effective potential $V_\mathrm{eff}$ (Eq.~\eqref{eq:Veff}) at $T=0$ and the thermal free energy landscape $F=U-TS$ at sufficient high $T$. At $T<T_c$, the minima of $V_\mathrm{neq}$ are at fan states. As $T$ increases, the minima at fan states become shallower and eventually become degenerate with the Y states. At this point, the overall magnitude of $V_\mathrm{neq}$ is also quite small. Above $T_c$, the Y states are true minima and grow deeper with increasing $T$. Note that this picture suggests coexisting peaks at both fan states and Y states at $T_c$ in the histogram. However, we do not observe such coexisting peaks in simulation. This is likely due to the fact that the peaks are too small comparing to the sampling noise.

Although a systematic classification of non-equilibrium phase transition is lacking, the above picture suggests that the transition from the fan phase to the Y phase resembles a weakly first order transition. It is also analogous to the transition in XY-clock model when the clock anisotropy changes sign. Let $\Phi$ stand for the order parameter angle of the XY model. Consider the clock anisotropy potential $\Delta = -g_6\cos(6\Phi)-g_{12}\cos(12\Phi)$. Tuning $g_6$ from positive to negative whist keeping $g_{12}>0$, the minima of $\Delta$ jump from $\Phi = m\pi/3$ to $\Phi = \pi/6+m\pi/3$, $m=0,1,2,\cdots5$. At the transition $g_6 = 0$, all twelve states are degenerate minima of $\Delta$.

The phase transitions at $B_0/J=3$ and $7$ can be analyzed in the same vein. It is advantageous to first consider $B_0/J=7$. In this case, we can also use the order parameter $\zeta_3$ to distinguish the fan phase ($\zeta_3 \to 1$) from the 2:1 phase ($\zeta_3 \to -1$). While $\zeta_3$ clearly shows a transition from the fan phase to the 2:1 phase as the temperature $T$ increases (Fig.~\ref{fig:finiteT}c$_3$), the manner in which the transition occurs is markedly different from $B_0/J=1$. At $T=1.2\times10^{-2}$, the histogram of $3\mathrm{arg}\Psi_x$ has a broad peak at $0$ corresponding to fan states (Fig.~\ref{fig:finiteT}(d$_3$)). When $T$ increases, the peak splits into two peaks. As $T$ increases further, these two peaks approach each other and eventually merge at $\pi$, which corresponds to the 2:1 states. The histogram of $\phi_\mathrm{A,B,C}$ mirrors the same process (Fig.~\ref{fig:finiteT}(e$_3$)). 

We interpret our results in terms of the non-equilibrium effective energy landscape $V_\mathrm{neq}$ as follows (Fig.~\ref{fig:vneq}(c)): at low $T$, $V_\mathrm{neq}$ has six degenerate minima at fan states. Increasing $T$ splits each of the minima into two, producing in total twelve degenerate minima. These minima eventually merge at the six 2:1 states. In other words, there is an intermediate phase that separates the fan phase at low temperature and the Y phase at high temperature. Crucially, this picture suggests that we can no longer interpret the crossing point of $\zeta_3$ as a single $T_c$. Instead, there are two separate transitions at $T_{c1,c2}$, corresponding to the onset and the end point of the intermediate phase, respectively. We shall return to this point in Sec.~\ref{sec:discuss}.

Similar to the case with $B_0/J=1$, we can make an analogy with the XY-clock model. Consider the clock anisotropy potential $\Delta = -g_6\cos(6\Phi)-g_{12}\cos(12\Phi)$. Tuning $g_6$ across 0 whilst keeping $g_{12}<0$, the six degenerate minima of $\Delta$, initially at $m\pi/3$ for large positive $g_6$, split into twelve minima, and then merge at $\pi/6+m\pi/3$. We note the same kind of physics arises in the context of height model as well~\cite{Ralko:2008}.

At $B_0/J=3$, yet another behavior emerges at the transition from the fan phase to the UUD phase. The order parameter $\zeta_3$ jumps rather abruptly from $1$ (fan phase) to $-1$ (UUD phase) as $T$ increases (Fig.~\ref{fig:finiteT}(c$_2$)). The peak of the $3\mathrm{arg}\Psi_x$ histogram (Fig.~\ref{fig:finiteT}(d$_2$)) is initially at $0$ (fan states) at low $T$. As $T$ increases, it splits into two peaks and then both approach $\pi$ (UUD states). Meanwhile, a satellite peak develops at $\pi$, which grows and eventually becomes the dominate peak. The presence of a satellite peak suggests that $V_\mathrm{neq}$ develops metastable minima at the UUD states, which eventually become global minima at higher $T$ (Fig.~\ref{fig:vneq}(b)). In other words, the transition from the fan phase to the UUD phase resembles a strongly first order transition. Using the crossing points of $\zeta_3$, we estimate $T_c \approx 3.26\times10^{-3}J$.

The transition occurs at much lower temperature for $B_0/J=3$ compared to $B_0/J=1$. This is due to the large entropy difference between the UUD states and the fan states, which is $\Delta S = 0.12k_\mathrm{B}$ per site. For comparison, the entropy difference between the fan states and the Y states $\Delta S=2.6\times10^{-3}k_\mathrm{B}$ per site at $B_0/J=1$, which is two orders of magnitude smaller. We have argued that, at the transition, $T_c\Delta S\sim B^2/J$. As a result, with the same driving strength, the transition at $B_0/J=3$ occurs at a temperature scale that is significantly lower than $B_0/J=1$.

\section{\label{sec:discuss} Discussion}

To summarize, we have shown the dynamical stabilization of the fan phase by periodic driving in the triangular XY antiferromagnet. As a result of the competition between the periodic driving and the thermal fluctuations, the late time steady state exhibits a temperature-driven non-equilibrium phase transition out of the fan phase. These results could be potentially tested in easy-plane triangular antiferromagnets. For instance, when the system is subject to a static field $B_0/J=1$ and an AC field $B_1/J = 0.025$ with frequency $\Omega/\sqrt{J/I} = 2.6$, the transition temperature from the fan phase to the Y phase $T_c/J = 0.03$. Using parameters from RbFe(MoO$_4$)$_2$, we estimate $B_0\approx 2\mathrm{T}$, $B_1\approx 50\mathrm{mT}$, $\Omega \approx 135\mathrm{GHz}$, and $T_c \approx 228\mathrm{mK}$. The sub-THz AC magnetic field is within the reach of current THz technology~\cite{Hayashi:2014}. We caution that a detailed modeling of the material is necessary to obtain a more accurate estimate.

The transitions from the fan phase to the other three thermal phases (Y, UUD, and 2:1) bear a resemblance to the transitions observed in the triangular XY antiferromagnet with quenched disorder~\cite{Maryasin:2013,Smirnov:2017}. In the latter model, the quenched disorder stabilize the fan phase through the order by quenched disorder mechanism~\cite{Henley:1989}, whereas the thermal fluctuations stabilize the other three. One therefore would expect temperature-driven transitions out of the fan phase. However, the transitions observed in our model, while resembling thermal phase transitions, are inherently out of equilibrium. Moreover, the presence of the quenched disorder in the latter model may have a significant impact on the nature of phase transition, whereas the quenched disorder is absent in our model.

Our work leaves a few interesting open questions. We observe from simulation that the transition from the fan phase to the 2:1 phase at $B_0/J=7$ is not direct. Instead, the system enters an intermediate phase that interpolates between the fan and the Y phases as the temperature $T$ increases. As the crossing point analysis in Sec.~\ref{sec:finiteT} is not directly applicable near the onset and end point temperature of the intermediate phase, a new analysis method is needed to locate them reliably. Furthermore, the system's rich behavior near the phase transitions calls for a thorough analytic treatment. Our analysis given in Sec.~\ref{sec:zeroT} completely discards the thermal fluctuations. It is therefore unable to capture the phase transition. Instead, in Sec.~\ref{sec:finiteT}, we invoke the concept of non-equilibrium effective energy landscape $V_\mathrm{neq}$ to interpret the simulation data. While intuitive, our picture should be put on a rigorous ground. Finally, it would be interesting to examine to what extent the results for the classical model carry over to the quantum XY model~\cite{Chubokov:1991} or the Heisenberg model~\cite{Kawamura:1985,Chubokov:1991,Gvozdikova:2011,Seabra:2011}. 

Looking beyond the triangular antiferromagnets, we think the non-equilibrium selection mechanism unveiled in Sec.~\ref{sec:zeroT} may be applicable to other classical frustrated systems that possess a continuously degeneracy ground state manifold. So long as the stiffness of the optically active normal modes depends on the ground state, driving these modes with AC magnetic field produces an effective potential that resembles Eq.~\eqref{eq:Veff}, which may dynamically stabilize ground states that are thermally unstable. It then follows that coupling the system to a thermal bath would result in non-equilibrium phase transitions similar to what we have found in Sec.~\ref{sec:finiteT}. In short, we believe that our work merely uncovers a corner of a potentially rich research direction.

\begin{acknowledgments}
We thank John Chalker and Collin Broholm for illuminating discussions. The work at the University of Oxford was supported in part by EPSRC Grant No. EP/N01930X/1. The work at MPI-PKS was supported in part by DFG under grant SFB 1143.
\end{acknowledgments}

\appendix

\section{\label{app:arclength} Arc-length parametrization}

In this section, we explain in detail the arc-length parametrization of the ground state manifold of the model Eq.~\eqref{eq:model}. In a ground state, the spins belonging to the same sublattice takes the same orientation. Let $\phi_\alpha$ denote the polar angle of the XY spins in sublattice $\alpha$, where $\alpha$ runs over the three sublattice labels A, B, and C. The energy of Eq.~\eqref{eq:model} is minimized if the following conditions are fulfilled:
\begin{align}
\sum_\alpha\cos\phi_\alpha = \frac{B_0}{3J};\quad
\sum_\alpha\sin\phi_\alpha = 0.
\end{align}
$\phi_\alpha$ span a three-dimensional torus $T^3$. The solution space of the above equations, or equivalently the ground state manifold, is one-dimensional, which may be viewed as a one-dimensional curve embedded in $T^3$. We thus may parametrize the solutions as $\phi_\alpha(s)$, where $s$ is the arc length parameter we introduced in the main text.

We next look for the functional form of $\phi_\alpha(s)$. To this end, we take derivatives of the ground state conditions with respect to $s$:
\begin{align}
\sum_{\alpha}\sin\phi_\alpha\dot{\phi}_\alpha = 0;\quad
\sum_{\alpha}\cos\phi_\alpha\dot{\phi}_\alpha = 0.
\end{align}
Recast the above in a more suggestive form:
\begin{align}
\mathbf{v}_{1}\cdot \dot{\bm{\phi}} = 0;\quad
\mathbf{v}_{2}\cdot \dot{\bm{\phi}} = 0.
\end{align}
Here, $\dot{\bm{\phi}}$ is the three-dimensional vector made of $\dot{\phi}_\alpha$. As $s$ is the arc-length of the ground state curve, $\dot{\bm{\phi}}$ is the tangent vector of the curve. In particular, $|\dot{\bm{\phi}}| = 1$. Likewise, $\mathbf{v}_1$ is a three-dimensional vector made of $\sin\phi_\alpha$, and $\mathbf{v}_2$ is made of $\cos\phi_\alpha$. The above equations show that $\dot{\bm{\phi}}\perp\mathbf{v}_{1,2}$, which is sufficient to determine $\dot{\bm{\phi}}$:
\begin{align}
\dot{\bm{\phi}} = \frac{\mathbf{v}_1\times\mathbf{v}_2}{|\mathbf{v}_1\times\mathbf{v}_2|}.
\end{align}
This is an autonomous system of first order differential equations, which fully determines $\phi_\alpha(s)$. Solving it numerically yields the ground state manifold visualized in Fig.~\ref{fig:zeroT}, column a.

\section{\label{app:derivation} Derivation of the effective potential}

In this section, we derive the effective potential Eq.~\eqref{eq:Veff} in more detail.  The starting point of the derivation is the Lagrangian Eq.~\eqref{eq:lagrangian}. We substitute the variational ansatz Eq.~\eqref{eq:ansatz} in to \eqref{eq:lagrangian}, expand it to the quadratic order in $A_\alpha$, and average over a time period of $2\pi/\Omega$~\cite{Whitham:1974}. We thus find the averaged Lagrangian:
\begin{align}
\overline{L} = \overline{K}-\overline{V}.
\end{align}
The averaged kinetic energy is given by,
\begin{align}
\overline{K} = &\frac{NI}{6}\dot{s}^2+\frac{NI}{12}\sum_\alpha[\dot{A}^{\phantom{\ast}}_\alpha\dot{A}^\ast_\alpha+\Omega^2 A^{\phantom{\ast}}_\alpha A^\ast_\alpha \nonumber\\
& +i\Omega(A^{\phantom{\ast}}_\alpha\dot{A}^\ast_\alpha-\dot{A}^{\phantom{\ast}}_\alpha A^\ast_\alpha)].
\end{align}
Here, the first term is the kinetic energy associated with the drifting motion in the ground state space, whereas the second term is the kinetic energy associated with the oscillation in the normal modes. The averaged potential energy is given by,
\begin{subequations}
\begin{align}
\overline{V} = \frac{N}{12}\sum_{\alpha}K_{\alpha\beta}A^\ast_{\alpha}A^{\phantom{\ast}}_\beta-\frac{NB_1}{6}\sum_{\alpha}m^y_\alpha \mathrm{Re}A^{\phantom{\ast}}_\alpha,
\end{align}
where $K_{\alpha\beta}$ is a $3\times3$ stiffness matrix:
\begin{align}
K_{\alpha\beta} = 3J\cos(\phi^{(0)}_\alpha-\phi^{(0)}_\beta).
\end{align}
$K_{\alpha\beta}$ depends on the ground state coordinate $s$ through $\phi^{(0)}_\alpha$, the spin polar angles in a ground state. The $3\times1$ vector $m^y_\alpha$ describes the coupling with the driving field:
\begin{align}
m^y_{\alpha} = \cos\phi^{(0)}_\alpha.
\end{align}
\end{subequations}
The averaged Rayleigh dissipation function is given by
\begin{align}
\overline{R} = &\frac{Nk}{6}\dot{s}^2+\frac{Nk}{12}\sum_\alpha[\dot{A}^{\phantom{\ast}}_\alpha\dot{A}^\ast_\alpha+\Omega^2 A^{\phantom{\ast}}_\alpha A^\ast_\alpha \nonumber\\
& +i\Omega(A^{\phantom{\ast}}_\alpha\dot{A}^\ast_\alpha-\dot{A}^{\phantom{\ast}}_\alpha A^\ast_\alpha)].
\end{align}

Using the Euler-Lagrangian-Rayleigh equation for $A_\alpha$ and $s$ yields the following equations of motion:
\begin{align}\label{eq:app:eom_A}
(ik\Omega-I\Omega^2)A_\alpha+\sum_{\beta}K_{\alpha\beta}A_\beta = B_1m^y_\alpha,
\end{align}
and
\begin{align}\label{eq:app:eom_s}
\frac{NI}{3}\ddot{s}+\frac{Nk}{3}\dot{s} = -N\frac{\partial V_\mathrm{eff}}{\partial s},
\end{align}
where the effective potential $V_\mathrm{eff}$ is defined through its derivative:
\begin{align}\label{eq:app:dvds}
\frac{\partial V_\mathrm{eff}}{\partial s} \equiv \frac{1}{12}\sum_{\alpha\beta} \frac{\partial K_{\alpha\beta}}{\partial s}A^\ast_\alpha A^{\phantom{\ast}}_\beta - \frac{B_1}{6}\sum_\alpha\frac{\partial m^y_\alpha}{\partial s} \mathrm{Re}A_\alpha.
\end{align}
In deriving Eq.~\eqref{eq:app:eom_A}, we have omitted all time derivatives of $A_\alpha$ thanks to the assumption $\dot{A}_\alpha \ll\Omega A_\alpha$.

We solve Eq.~\eqref{eq:app:eom_A} for $A_\alpha$ and plug it into Eq.~\eqref{eq:app:dvds}:
\begin{align}
\frac{\partial V_\mathrm{eff}}{\partial s} = &\frac{B^2_1}{12I}\sum_{\lambda} \left[\frac{(m^y_\lambda)^2}{(\omega^2_\lambda-\Omega^2)^2+k^2\Omega^2/I^2}\frac{\partial \omega^2_\lambda}{\partial s}\right.\nonumber\\
& \left.-\frac{\partial (m^y_\lambda)^2}{\partial s}\frac{\omega^2_\lambda-\Omega^2}{(\omega^2_\lambda-\Omega^2)^2+k^2\Omega^2/I^2}\right].
\end{align}
Here, the summation is over the two normal modes $\lambda$. $m^y_\lambda$ and $\omega_\lambda$ are respectively the $y$-dipole moment and the frequency of the normal mode $\lambda$. In the weak damping limit, $V_\mathrm{eff}$ can be integrated approximately:
\begin{align}
V_\mathrm{eff} = \frac{B^2_1}{12I}\sum_\lambda\frac{\Omega^2-\omega^2_\lambda}{(\omega^2_\lambda-\Omega^2)^2+k^2\Omega^2/I^2}(m^y_\lambda)^2,
\end{align}
which is the result given in Eq.~\eqref{eq:Veff}. The error is of order $k^2/IJ$.

\section{\label{app:algorithm} Integrating the Langevin equation}

In this section, we provide the details of the numerical procedure for integrating the Langevin equation Eq.~\eqref{eq:langevin}. We employ the Bussi-Parrinello algorithm~\cite{Bussi:2007}. In this algorithm, each step of the evolution is split into four stages. In the second and the third stages, the system decouples from the bath and evolves according to its Hamiltonian. In the first and the last stages, the system equilibrates with the bath. The explicit formulae at the four stages are given by:
\begin{subequations}\label{eq:BP}
\begin{align}
L_i(t^+) &= c_1 L_i(t)+c_2 R_i(t);\\
\phi_i(t+\Delta t) &= \phi_i(t)+L_i(t)\Delta t + \frac{\tau_i(t)}{2}\Delta t^2;\\
L_i(t^-+\Delta t) &= L_i(t^+)+\frac{\tau_i(t)+\tau_i(t+\Delta t)}{2}\Delta t;\\
L_i(t+\Delta t) &=c_1 L_i(t^-+\Delta t)+c_2 R_i(t+\Delta t).
\end{align}
\end{subequations}
Here, $L_i$ and $\phi_i$ are respectively the angular velocity and the polar angle of the XY spin. $\tau_i$ is the deterministic torque. For the sake of simplicity, we have rescaled time and energy such that the spin's rotational inertia $I\to 1$ and exchange constant $J\to 1$. $R_i(t)$ and $R_i(t+\Delta t)$ are pseudo-random numbers drawn from a standard normal distribution. The dimensionless constants $c_1 = \exp(-k\Delta t/2)$ and $c_2 = \sqrt{(1-c^2_1)k_BT}$, where $k$ and $k_BT$ are respectively the dimensionless damping constant and temperature. In particular, if we set the bath temperature $k_BT=0$, $c_2 = 0$, Eq.~\eqref{eq:BP}  reduces to the classic velocity Verlet algorithm.

In the simulation, we set the step width $\Delta t$ to be about $1/200$ of the driving period, i.e. $0.01\pi/\Omega$. Convergence is checked by reducing $\Delta t$. Each run starts with an initial configuration of $\phi_i$ and $L_i$ drawn from a Monte Carlo simulation. We equilibrate the system by running the algorithm for $10^5$ steps before the driving field is turned on. After the system reaches the late time steady state, we record the system's configurations at stroboscopic times $t_i$ when the periodic driving field vanishes instantaneously, i.e. $B_1(t_i)=0$. We collect more than $10^5$ samples in each run, and run 8 times with different initial conditions for each model parameter.

\bibliography{tri_drive.bib}

\begin{thebibliography}{46}%
\makeatletter
\providecommand \@ifxundefined [1]{%
 \@ifx{#1\undefined}
}%
\providecommand \@ifnum [1]{%
 \ifnum #1\expandafter \@firstoftwo
 \else \expandafter \@secondoftwo
 \fi
}%
\providecommand \@ifx [1]{%
 \ifx #1\expandafter \@firstoftwo
 \else \expandafter \@secondoftwo
 \fi
}%
\providecommand \natexlab [1]{#1}%
\providecommand \enquote  [1]{``#1''}%
\providecommand \bibnamefont  [1]{#1}%
\providecommand \bibfnamefont [1]{#1}%
\providecommand \citenamefont [1]{#1}%
\providecommand \href@noop [0]{\@secondoftwo}%
\providecommand \href [0]{\begingroup \@sanitize@url \@href}%
\providecommand \@href[1]{\@@startlink{#1}\@@href}%
\providecommand \@@href[1]{\endgroup#1\@@endlink}%
\providecommand \@sanitize@url [0]{\catcode `\\12\catcode `\$12\catcode
  `\&12\catcode `\#12\catcode `\^12\catcode `\_12\catcode `\%12\relax}%
\providecommand \@@startlink[1]{}%
\providecommand \@@endlink[0]{}%
\providecommand \url  [0]{\begingroup\@sanitize@url \@url }%
\providecommand \@url [1]{\endgroup\@href {#1}{\urlprefix }}%
\providecommand \urlprefix  [0]{URL }%
\providecommand \Eprint [0]{\href }%
\providecommand \doibase [0]{http://dx.doi.org/}%
\providecommand \selectlanguage [0]{\@gobble}%
\providecommand \bibinfo  [0]{\@secondoftwo}%
\providecommand \bibfield  [0]{\@secondoftwo}%
\providecommand \translation [1]{[#1]}%
\providecommand \BibitemOpen [0]{}%
\providecommand \bibitemStop [0]{}%
\providecommand \bibitemNoStop [0]{.\EOS\space}%
\providecommand \EOS [0]{\spacefactor3000\relax}%
\providecommand \BibitemShut  [1]{\csname bibitem#1\endcsname}%
\let\auto@bib@innerbib\@empty
%</preamble>
\bibitem [{\citenamefont {B{\'e}nard}(1901)}]{Benard:1901}%
  \BibitemOpen
  \bibfield  {author} {\bibinfo {author} {\bibfnamefont {H.}~\bibnamefont
  {B{\'e}nard}},\ }\href {\doibase 10.1051/jphystap:0190100100025400}
  {\bibfield  {journal} {\bibinfo  {journal} {J. Phys. Theor. Appl.}\ }\textbf
  {\bibinfo {volume} {10}},\ \bibinfo {pages} {254} (\bibinfo {year}
  {1901})}\BibitemShut {NoStop}%
\bibitem [{\citenamefont {Dou{\c c}ot}\ and\ \citenamefont
  {Simon}(1998)}]{Doucot:1998}%
  \BibitemOpen
  \bibfield  {author} {\bibinfo {author} {\bibfnamefont {B.}~\bibnamefont
  {Dou{\c c}ot}}\ and\ \bibinfo {author} {\bibfnamefont {P.}~\bibnamefont
  {Simon}},\ }\href {http://stacks.iop.org/0305-4470/31/i=28/a=005} {\bibfield
  {journal} {\bibinfo  {journal} {Journal of Physics A: Mathematical and
  General}\ }\textbf {\bibinfo {volume} {31}},\ \bibinfo {pages} {5855}
  (\bibinfo {year} {1998})}\BibitemShut {NoStop}%
\bibitem [{\citenamefont {Wan}\ and\ \citenamefont
  {Moessner}(2017)}]{Wan:2017}%
  \BibitemOpen
  \bibfield  {author} {\bibinfo {author} {\bibfnamefont {Y.}~\bibnamefont
  {Wan}}\ and\ \bibinfo {author} {\bibfnamefont {R.}~\bibnamefont {Moessner}},\
  }\href {\doibase 10.1103/PhysRevLett.119.167203} {\bibfield  {journal}
  {\bibinfo  {journal} {Phys. Rev. Lett.}\ }\textbf {\bibinfo {volume} {119}},\
  \bibinfo {pages} {167203} (\bibinfo {year} {2017})}\BibitemShut {NoStop}%
\bibitem [{\citenamefont {Villain}\ \emph {et~al.}(1980)\citenamefont
  {Villain}, \citenamefont {Bidaux}, \citenamefont {Carton},\ and\
  \citenamefont {Conte}}]{Villain:1980}%
  \BibitemOpen
  \bibfield  {author} {\bibinfo {author} {\bibfnamefont {J.}~\bibnamefont
  {Villain}}, \bibinfo {author} {\bibfnamefont {R.}~\bibnamefont {Bidaux}},
  \bibinfo {author} {\bibfnamefont {J.-P.}\ \bibnamefont {Carton}}, \ and\
  \bibinfo {author} {\bibfnamefont {R.}~\bibnamefont {Conte}},\ }\href@noop {}
  {\bibfield  {journal} {\bibinfo  {journal} {J. Phys. France}\ }\textbf
  {\bibinfo {volume} {41}},\ \bibinfo {pages} {1263} (\bibinfo {year}
  {1980})}\BibitemShut {NoStop}%
\bibitem [{\citenamefont {Shender}(1982)}]{Shender:1982}%
  \BibitemOpen
  \bibfield  {author} {\bibinfo {author} {\bibfnamefont {E.}~\bibnamefont
  {Shender}},\ }\href@noop {} {\bibfield  {journal} {\bibinfo  {journal} {Zh.
  Eksp. Teor. Fiz}\ }\textbf {\bibinfo {volume} {83}},\ \bibinfo {pages} {326}
  (\bibinfo {year} {1982})}\BibitemShut {NoStop}%
\bibitem [{\citenamefont {Kawamura}(1984)}]{Kawamura:1984}%
  \BibitemOpen
  \bibfield  {author} {\bibinfo {author} {\bibfnamefont {H.}~\bibnamefont
  {Kawamura}},\ }\href {\doibase 10.1143/JPSJ.53.2452} {\bibfield  {journal}
  {\bibinfo  {journal} {Journal of the Physical Society of Japan}\ }\textbf
  {\bibinfo {volume} {53}},\ \bibinfo {pages} {2452} (\bibinfo {year}
  {1984})}\BibitemShut {NoStop}%
\bibitem [{\citenamefont {Henley}(1989)}]{Henley:1989}%
  \BibitemOpen
  \bibfield  {author} {\bibinfo {author} {\bibfnamefont {C.~L.}\ \bibnamefont
  {Henley}},\ }\href {\doibase 10.1103/PhysRevLett.62.2056} {\bibfield
  {journal} {\bibinfo  {journal} {Phys. Rev. Lett.}\ }\textbf {\bibinfo
  {volume} {62}},\ \bibinfo {pages} {2056} (\bibinfo {year}
  {1989})}\BibitemShut {NoStop}%
\bibitem [{\citenamefont {Moessner}\ and\ \citenamefont
  {Chalker}(1998)}]{Moessner:1998}%
  \BibitemOpen
  \bibfield  {author} {\bibinfo {author} {\bibfnamefont {R.}~\bibnamefont
  {Moessner}}\ and\ \bibinfo {author} {\bibfnamefont {J.~T.}\ \bibnamefont
  {Chalker}},\ }\href {\doibase 10.1103/PhysRevB.58.12049} {\bibfield
  {journal} {\bibinfo  {journal} {Phys. Rev. B}\ }\textbf {\bibinfo {volume}
  {58}},\ \bibinfo {pages} {12049} (\bibinfo {year} {1998})}\BibitemShut
  {NoStop}%
\bibitem [{\citenamefont {Kapitsa}(1951)}]{Kapitsa:1951}%
  \BibitemOpen
  \bibfield  {author} {\bibinfo {author} {\bibfnamefont {P.~L.}\ \bibnamefont
  {Kapitsa}},\ }\href {\doibase 10.3367/UFNr.0044.195105b.0007} {\bibfield
  {journal} {\bibinfo  {journal} {Usp. Fiz. Nauk}\ }\textbf {\bibinfo {volume}
  {44}},\ \bibinfo {pages} {7} (\bibinfo {year} {1951})}\BibitemShut {NoStop}%
\bibitem [{\citenamefont {Landau}\ and\ \citenamefont
  {Lifshitz}(1976)}]{Landau:1976}%
  \BibitemOpen
  \bibfield  {author} {\bibinfo {author} {\bibfnamefont {L.~D.}\ \bibnamefont
  {Landau}}\ and\ \bibinfo {author} {\bibfnamefont {E.~M.}\ \bibnamefont
  {Lifshitz}},\ }\href@noop {} {\emph {\bibinfo {title} {Mechanics}}},\
  \bibinfo {edition} {3rd}\ ed.,\ \bibinfo {series} {Course of Theoretical
  Physics}, Vol.~\bibinfo {volume} {1}\ (\bibinfo  {publisher}
  {Butterworth-Heinemann},\ \bibinfo {year} {1976})\BibitemShut {NoStop}%
\bibitem [{\citenamefont {Grifoni}\ and\ \citenamefont
  {H{\"a}nggi}(1998)}]{Grifoni:1998}%
  \BibitemOpen
  \bibfield  {author} {\bibinfo {author} {\bibfnamefont {M.}~\bibnamefont
  {Grifoni}}\ and\ \bibinfo {author} {\bibfnamefont {P.}~\bibnamefont
  {H{\"a}nggi}},\ }\href@noop {} {\bibfield  {journal} {\bibinfo  {journal}
  {Physics Reports}\ }\textbf {\bibinfo {volume} {304}},\ \bibinfo {pages} {229
  } (\bibinfo {year} {1998})}\BibitemShut {NoStop}%
\bibitem [{\citenamefont {Bukov}\ \emph {et~al.}(2015)\citenamefont {Bukov},
  \citenamefont {D'Alessio},\ and\ \citenamefont {Polkovnikov}}]{Bukov:2015}%
  \BibitemOpen
  \bibfield  {author} {\bibinfo {author} {\bibfnamefont {M.}~\bibnamefont
  {Bukov}}, \bibinfo {author} {\bibfnamefont {L.}~\bibnamefont {D'Alessio}}, \
  and\ \bibinfo {author} {\bibfnamefont {A.}~\bibnamefont {Polkovnikov}},\
  }\href {\doibase 10.1080/00018732.2015.1055918} {\bibfield  {journal}
  {\bibinfo  {journal} {Advances in Physics}\ }\textbf {\bibinfo {volume}
  {64}},\ \bibinfo {pages} {139} (\bibinfo {year} {2015})}\BibitemShut
  {NoStop}%
\bibitem [{\citenamefont {{Oka}}\ and\ \citenamefont
  {{Kitamura}}(2018)}]{Oka:2018}%
  \BibitemOpen
  \bibfield  {author} {\bibinfo {author} {\bibfnamefont {T.}~\bibnamefont
  {{Oka}}}\ and\ \bibinfo {author} {\bibfnamefont {S.}~\bibnamefont
  {{Kitamura}}},\ }\href@noop {} {\bibfield  {journal} {\bibinfo  {journal}
  {ArXiv e-prints}\ } (\bibinfo {year} {2018})},\ \Eprint
  {http://arxiv.org/abs/1804.03212} {arXiv:1804.03212 [cond-mat.str-el]}
  \BibitemShut {NoStop}%
\bibitem [{\citenamefont {Mitamura}\ \emph {et~al.}(2014)\citenamefont
  {Mitamura}, \citenamefont {Watanuki}, \citenamefont {Kaneko}, \citenamefont
  {Onozaki}, \citenamefont {Amou}, \citenamefont {Kittaka}, \citenamefont
  {Kobayashi}, \citenamefont {Shimura}, \citenamefont {Yamamoto}, \citenamefont
  {Suzuki}, \citenamefont {Chi},\ and\ \citenamefont
  {Sakakibara}}]{Mitamura:2014}%
  \BibitemOpen
  \bibfield  {author} {\bibinfo {author} {\bibfnamefont {H.}~\bibnamefont
  {Mitamura}}, \bibinfo {author} {\bibfnamefont {R.}~\bibnamefont {Watanuki}},
  \bibinfo {author} {\bibfnamefont {K.}~\bibnamefont {Kaneko}}, \bibinfo
  {author} {\bibfnamefont {N.}~\bibnamefont {Onozaki}}, \bibinfo {author}
  {\bibfnamefont {Y.}~\bibnamefont {Amou}}, \bibinfo {author} {\bibfnamefont
  {S.}~\bibnamefont {Kittaka}}, \bibinfo {author} {\bibfnamefont
  {R.}~\bibnamefont {Kobayashi}}, \bibinfo {author} {\bibfnamefont
  {Y.}~\bibnamefont {Shimura}}, \bibinfo {author} {\bibfnamefont
  {I.}~\bibnamefont {Yamamoto}}, \bibinfo {author} {\bibfnamefont
  {K.}~\bibnamefont {Suzuki}}, \bibinfo {author} {\bibfnamefont
  {S.}~\bibnamefont {Chi}}, \ and\ \bibinfo {author} {\bibfnamefont
  {T.}~\bibnamefont {Sakakibara}},\ }\href {\doibase
  10.1103/PhysRevLett.113.147202} {\bibfield  {journal} {\bibinfo  {journal}
  {Phys. Rev. Lett.}\ }\textbf {\bibinfo {volume} {113}},\ \bibinfo {pages}
  {147202} (\bibinfo {year} {2014})}\BibitemShut {NoStop}%
\bibitem [{\citenamefont {Takayoshi}\ \emph {et~al.}(2014)\citenamefont
  {Takayoshi}, \citenamefont {Aoki},\ and\ \citenamefont
  {Oka}}]{Takayoshi:2014}%
  \BibitemOpen
  \bibfield  {author} {\bibinfo {author} {\bibfnamefont {S.}~\bibnamefont
  {Takayoshi}}, \bibinfo {author} {\bibfnamefont {H.}~\bibnamefont {Aoki}}, \
  and\ \bibinfo {author} {\bibfnamefont {T.}~\bibnamefont {Oka}},\ }\href
  {\doibase 10.1103/PhysRevB.90.085150} {\bibfield  {journal} {\bibinfo
  {journal} {Phys. Rev. B}\ }\textbf {\bibinfo {volume} {90}},\ \bibinfo
  {pages} {085150} (\bibinfo {year} {2014})}\BibitemShut {NoStop}%
\bibitem [{\citenamefont {Sato}\ \emph {et~al.}(2016)\citenamefont {Sato},
  \citenamefont {Takayoshi},\ and\ \citenamefont {Oka}}]{Sato:2016}%
  \BibitemOpen
  \bibfield  {author} {\bibinfo {author} {\bibfnamefont {M.}~\bibnamefont
  {Sato}}, \bibinfo {author} {\bibfnamefont {S.}~\bibnamefont {Takayoshi}}, \
  and\ \bibinfo {author} {\bibfnamefont {T.}~\bibnamefont {Oka}},\ }\href
  {\doibase 10.1103/PhysRevLett.117.147202} {\bibfield  {journal} {\bibinfo
  {journal} {Phys. Rev. Lett.}\ }\textbf {\bibinfo {volume} {117}},\ \bibinfo
  {pages} {147202} (\bibinfo {year} {2016})}\BibitemShut {NoStop}%
\bibitem [{\citenamefont {Kitamura}\ \emph {et~al.}(2017)\citenamefont
  {Kitamura}, \citenamefont {Oka},\ and\ \citenamefont {Aoki}}]{Kitamura:2017}%
  \BibitemOpen
  \bibfield  {author} {\bibinfo {author} {\bibfnamefont {S.}~\bibnamefont
  {Kitamura}}, \bibinfo {author} {\bibfnamefont {T.}~\bibnamefont {Oka}}, \
  and\ \bibinfo {author} {\bibfnamefont {H.}~\bibnamefont {Aoki}},\ }\href
  {\doibase 10.1103/PhysRevB.96.014406} {\bibfield  {journal} {\bibinfo
  {journal} {Phys. Rev. B}\ }\textbf {\bibinfo {volume} {96}},\ \bibinfo
  {pages} {014406} (\bibinfo {year} {2017})}\BibitemShut {NoStop}%
\bibitem [{\citenamefont {Claassen}\ \emph {et~al.}(2017)\citenamefont
  {Claassen}, \citenamefont {Jiang}, \citenamefont {Moritz},\ and\
  \citenamefont {Devereaux}}]{Claassen:2017}%
  \BibitemOpen
  \bibfield  {author} {\bibinfo {author} {\bibfnamefont {M.}~\bibnamefont
  {Claassen}}, \bibinfo {author} {\bibfnamefont {H.-C.}\ \bibnamefont {Jiang}},
  \bibinfo {author} {\bibfnamefont {B.}~\bibnamefont {Moritz}}, \ and\ \bibinfo
  {author} {\bibfnamefont {T.~P.}\ \bibnamefont {Devereaux}},\ }\href {\doibase
  10.1038/s41467-017-00876-y} {\bibfield  {journal} {\bibinfo  {journal}
  {Nature Communications}\ }\textbf {\bibinfo {volume} {8}},\ \bibinfo {pages}
  {1192} (\bibinfo {year} {2017})}\BibitemShut {NoStop}%
\bibitem [{\citenamefont {Lee}\ \emph {et~al.}(1984)\citenamefont {Lee},
  \citenamefont {Joannopoulos}, \citenamefont {Negele},\ and\ \citenamefont
  {Landau}}]{Lee:1984}%
  \BibitemOpen
  \bibfield  {author} {\bibinfo {author} {\bibfnamefont {D.~H.}\ \bibnamefont
  {Lee}}, \bibinfo {author} {\bibfnamefont {J.~D.}\ \bibnamefont
  {Joannopoulos}}, \bibinfo {author} {\bibfnamefont {J.~W.}\ \bibnamefont
  {Negele}}, \ and\ \bibinfo {author} {\bibfnamefont {D.~P.}\ \bibnamefont
  {Landau}},\ }\href {\doibase 10.1103/PhysRevLett.52.433} {\bibfield
  {journal} {\bibinfo  {journal} {Phys. Rev. Lett.}\ }\textbf {\bibinfo
  {volume} {52}},\ \bibinfo {pages} {433} (\bibinfo {year} {1984})}\BibitemShut
  {NoStop}%
\bibitem [{\citenamefont {Lee}\ \emph {et~al.}(1986)\citenamefont {Lee},
  \citenamefont {Joannopoulos}, \citenamefont {Negele},\ and\ \citenamefont
  {Landau}}]{Lee:1986}%
  \BibitemOpen
  \bibfield  {author} {\bibinfo {author} {\bibfnamefont {D.~H.}\ \bibnamefont
  {Lee}}, \bibinfo {author} {\bibfnamefont {J.~D.}\ \bibnamefont
  {Joannopoulos}}, \bibinfo {author} {\bibfnamefont {J.~W.}\ \bibnamefont
  {Negele}}, \ and\ \bibinfo {author} {\bibfnamefont {D.~P.}\ \bibnamefont
  {Landau}},\ }\href {\doibase 10.1103/PhysRevB.33.450} {\bibfield  {journal}
  {\bibinfo  {journal} {Phys. Rev. B}\ }\textbf {\bibinfo {volume} {33}},\
  \bibinfo {pages} {450} (\bibinfo {year} {1986})}\BibitemShut {NoStop}%
\bibitem [{\citenamefont {Chubokov}\ and\ \citenamefont
  {Golosov}(1991)}]{Chubokov:1991}%
  \BibitemOpen
  \bibfield  {author} {\bibinfo {author} {\bibfnamefont {A.~V.}\ \bibnamefont
  {Chubokov}}\ and\ \bibinfo {author} {\bibfnamefont {D.~I.}\ \bibnamefont
  {Golosov}},\ }\href {http://stacks.iop.org/0953-8984/3/i=1/a=005} {\bibfield
  {journal} {\bibinfo  {journal} {Journal of Physics: Condensed Matter}\
  }\textbf {\bibinfo {volume} {3}},\ \bibinfo {pages} {69} (\bibinfo {year}
  {1991})}\BibitemShut {NoStop}%
\bibitem [{\citenamefont {Collins}\ and\ \citenamefont
  {Petrenko}(1997)}]{Collins:1997}%
  \BibitemOpen
  \bibfield  {author} {\bibinfo {author} {\bibfnamefont {M.~F.}\ \bibnamefont
  {Collins}}\ and\ \bibinfo {author} {\bibfnamefont {O.~A.}\ \bibnamefont
  {Petrenko}},\ }\href {\doibase 10.1139/p97-007} {\bibfield  {journal}
  {\bibinfo  {journal} {Canadian Journal of Physics}\ }\textbf {\bibinfo
  {volume} {75}},\ \bibinfo {pages} {605} (\bibinfo {year} {1997})}\BibitemShut
  {NoStop}%
\bibitem [{\citenamefont {Inami}\ \emph {et~al.}(1996)\citenamefont {Inami},
  \citenamefont {Ajiro},\ and\ \citenamefont {Goto}}]{Inami:1996}%
  \BibitemOpen
  \bibfield  {author} {\bibinfo {author} {\bibfnamefont {T.}~\bibnamefont
  {Inami}}, \bibinfo {author} {\bibfnamefont {Y.}~\bibnamefont {Ajiro}}, \ and\
  \bibinfo {author} {\bibfnamefont {T.}~\bibnamefont {Goto}},\ }\href {\doibase
  10.1143/JPSJ.65.2374} {\bibfield  {journal} {\bibinfo  {journal} {Journal of
  the Physical Society of Japan}\ }\textbf {\bibinfo {volume} {65}},\ \bibinfo
  {pages} {2374} (\bibinfo {year} {1996})}\BibitemShut {NoStop}%
\bibitem [{\citenamefont {Svistov}\ \emph {et~al.}(2003)\citenamefont
  {Svistov}, \citenamefont {Smirnov}, \citenamefont {Prozorova}, \citenamefont
  {Petrenko}, \citenamefont {Demianets},\ and\ \citenamefont
  {Shapiro}}]{Svistov:2003}%
  \BibitemOpen
  \bibfield  {author} {\bibinfo {author} {\bibfnamefont {L.~E.}\ \bibnamefont
  {Svistov}}, \bibinfo {author} {\bibfnamefont {A.~I.}\ \bibnamefont
  {Smirnov}}, \bibinfo {author} {\bibfnamefont {L.~A.}\ \bibnamefont
  {Prozorova}}, \bibinfo {author} {\bibfnamefont {O.~A.}\ \bibnamefont
  {Petrenko}}, \bibinfo {author} {\bibfnamefont {L.~N.}\ \bibnamefont
  {Demianets}}, \ and\ \bibinfo {author} {\bibfnamefont {A.~Y.}\ \bibnamefont
  {Shapiro}},\ }\href {\doibase 10.1103/PhysRevB.67.094434} {\bibfield
  {journal} {\bibinfo  {journal} {Phys. Rev. B}\ }\textbf {\bibinfo {volume}
  {67}},\ \bibinfo {pages} {094434} (\bibinfo {year} {2003})}\BibitemShut
  {NoStop}%
\bibitem [{\citenamefont {Svistov}\ \emph {et~al.}(2006)\citenamefont
  {Svistov}, \citenamefont {Smirnov}, \citenamefont {Prozorova}, \citenamefont
  {Petrenko}, \citenamefont {Micheler}, \citenamefont {B\"uttgen},
  \citenamefont {Shapiro},\ and\ \citenamefont {Demianets}}]{Svistov:2006}%
  \BibitemOpen
  \bibfield  {author} {\bibinfo {author} {\bibfnamefont {L.~E.}\ \bibnamefont
  {Svistov}}, \bibinfo {author} {\bibfnamefont {A.~I.}\ \bibnamefont
  {Smirnov}}, \bibinfo {author} {\bibfnamefont {L.~A.}\ \bibnamefont
  {Prozorova}}, \bibinfo {author} {\bibfnamefont {O.~A.}\ \bibnamefont
  {Petrenko}}, \bibinfo {author} {\bibfnamefont {A.}~\bibnamefont {Micheler}},
  \bibinfo {author} {\bibfnamefont {N.}~\bibnamefont {B\"uttgen}}, \bibinfo
  {author} {\bibfnamefont {A.~Y.}\ \bibnamefont {Shapiro}}, \ and\ \bibinfo
  {author} {\bibfnamefont {L.~N.}\ \bibnamefont {Demianets}},\ }\href {\doibase
  10.1103/PhysRevB.74.024412} {\bibfield  {journal} {\bibinfo  {journal} {Phys.
  Rev. B}\ }\textbf {\bibinfo {volume} {74}},\ \bibinfo {pages} {024412}
  (\bibinfo {year} {2006})}\BibitemShut {NoStop}%
\bibitem [{\citenamefont {Kenzelmann}\ \emph {et~al.}(2007)\citenamefont
  {Kenzelmann}, \citenamefont {Lawes}, \citenamefont {Harris}, \citenamefont
  {Gasparovic}, \citenamefont {Broholm}, \citenamefont {Ramirez}, \citenamefont
  {Jorge}, \citenamefont {Jaime}, \citenamefont {Park}, \citenamefont {Huang},
  \citenamefont {Shapiro},\ and\ \citenamefont {Demianets}}]{Kenzelmann:2007}%
  \BibitemOpen
  \bibfield  {author} {\bibinfo {author} {\bibfnamefont {M.}~\bibnamefont
  {Kenzelmann}}, \bibinfo {author} {\bibfnamefont {G.}~\bibnamefont {Lawes}},
  \bibinfo {author} {\bibfnamefont {A.~B.}\ \bibnamefont {Harris}}, \bibinfo
  {author} {\bibfnamefont {G.}~\bibnamefont {Gasparovic}}, \bibinfo {author}
  {\bibfnamefont {C.}~\bibnamefont {Broholm}}, \bibinfo {author} {\bibfnamefont
  {A.~P.}\ \bibnamefont {Ramirez}}, \bibinfo {author} {\bibfnamefont {G.~A.}\
  \bibnamefont {Jorge}}, \bibinfo {author} {\bibfnamefont {M.}~\bibnamefont
  {Jaime}}, \bibinfo {author} {\bibfnamefont {S.}~\bibnamefont {Park}},
  \bibinfo {author} {\bibfnamefont {Q.}~\bibnamefont {Huang}}, \bibinfo
  {author} {\bibfnamefont {A.~Y.}\ \bibnamefont {Shapiro}}, \ and\ \bibinfo
  {author} {\bibfnamefont {L.~A.}\ \bibnamefont {Demianets}},\ }\href {\doibase
  10.1103/PhysRevLett.98.267205} {\bibfield  {journal} {\bibinfo  {journal}
  {Phys. Rev. Lett.}\ }\textbf {\bibinfo {volume} {98}},\ \bibinfo {pages}
  {267205} (\bibinfo {year} {2007})}\BibitemShut {NoStop}%
\bibitem [{\citenamefont {Smirnov}\ \emph {et~al.}(2007)\citenamefont
  {Smirnov}, \citenamefont {Yashiro}, \citenamefont {Kimura}, \citenamefont
  {Hagiwara}, \citenamefont {Narumi}, \citenamefont {Kindo}, \citenamefont
  {Kikkawa}, \citenamefont {Katsumata}, \citenamefont {Shapiro},\ and\
  \citenamefont {Demianets}}]{Smirnov:2007}%
  \BibitemOpen
  \bibfield  {author} {\bibinfo {author} {\bibfnamefont {A.~I.}\ \bibnamefont
  {Smirnov}}, \bibinfo {author} {\bibfnamefont {H.}~\bibnamefont {Yashiro}},
  \bibinfo {author} {\bibfnamefont {S.}~\bibnamefont {Kimura}}, \bibinfo
  {author} {\bibfnamefont {M.}~\bibnamefont {Hagiwara}}, \bibinfo {author}
  {\bibfnamefont {Y.}~\bibnamefont {Narumi}}, \bibinfo {author} {\bibfnamefont
  {K.}~\bibnamefont {Kindo}}, \bibinfo {author} {\bibfnamefont
  {A.}~\bibnamefont {Kikkawa}}, \bibinfo {author} {\bibfnamefont
  {K.}~\bibnamefont {Katsumata}}, \bibinfo {author} {\bibfnamefont {A.~Y.}\
  \bibnamefont {Shapiro}}, \ and\ \bibinfo {author} {\bibfnamefont {L.~N.}\
  \bibnamefont {Demianets}},\ }\href {\doibase 10.1103/PhysRevB.75.134412}
  {\bibfield  {journal} {\bibinfo  {journal} {Phys. Rev. B}\ }\textbf {\bibinfo
  {volume} {75}},\ \bibinfo {pages} {134412} (\bibinfo {year}
  {2007})}\BibitemShut {NoStop}%
\bibitem [{\citenamefont {Ribeiro}\ and\ \citenamefont
  {Perez-Mato}(2011)}]{Ribeiro:2011}%
  \BibitemOpen
  \bibfield  {author} {\bibinfo {author} {\bibfnamefont {J.~L.}\ \bibnamefont
  {Ribeiro}}\ and\ \bibinfo {author} {\bibfnamefont {J.~M.}\ \bibnamefont
  {Perez-Mato}},\ }\href {http://stacks.iop.org/0953-8984/23/i=44/a=446003}
  {\bibfield  {journal} {\bibinfo  {journal} {Journal of Physics: Condensed
  Matter}\ }\textbf {\bibinfo {volume} {23}},\ \bibinfo {pages} {446003}
  (\bibinfo {year} {2011})}\BibitemShut {NoStop}%
\bibitem [{\citenamefont {Hearmon}\ \emph {et~al.}(2012)\citenamefont
  {Hearmon}, \citenamefont {Fabrizi}, \citenamefont {Chapon}, \citenamefont
  {Johnson}, \citenamefont {Prabhakaran}, \citenamefont {Streltsov},
  \citenamefont {Brown},\ and\ \citenamefont {Radaelli}}]{Hearmon:2012}%
  \BibitemOpen
  \bibfield  {author} {\bibinfo {author} {\bibfnamefont {A.~J.}\ \bibnamefont
  {Hearmon}}, \bibinfo {author} {\bibfnamefont {F.}~\bibnamefont {Fabrizi}},
  \bibinfo {author} {\bibfnamefont {L.~C.}\ \bibnamefont {Chapon}}, \bibinfo
  {author} {\bibfnamefont {R.~D.}\ \bibnamefont {Johnson}}, \bibinfo {author}
  {\bibfnamefont {D.}~\bibnamefont {Prabhakaran}}, \bibinfo {author}
  {\bibfnamefont {S.~V.}\ \bibnamefont {Streltsov}}, \bibinfo {author}
  {\bibfnamefont {P.~J.}\ \bibnamefont {Brown}}, \ and\ \bibinfo {author}
  {\bibfnamefont {P.~G.}\ \bibnamefont {Radaelli}},\ }\href {\doibase
  10.1103/PhysRevLett.108.237201} {\bibfield  {journal} {\bibinfo  {journal}
  {Phys. Rev. Lett.}\ }\textbf {\bibinfo {volume} {108}},\ \bibinfo {pages}
  {237201} (\bibinfo {year} {2012})}\BibitemShut {NoStop}%
\bibitem [{\citenamefont {White}\ \emph {et~al.}(2013)\citenamefont {White},
  \citenamefont {Niedermayer}, \citenamefont {Gasparovic}, \citenamefont
  {Broholm}, \citenamefont {Park}, \citenamefont {Shapiro}, \citenamefont
  {Demianets},\ and\ \citenamefont {Kenzelmann}}]{White:2013}%
  \BibitemOpen
  \bibfield  {author} {\bibinfo {author} {\bibfnamefont {J.~S.}\ \bibnamefont
  {White}}, \bibinfo {author} {\bibfnamefont {C.}~\bibnamefont {Niedermayer}},
  \bibinfo {author} {\bibfnamefont {G.}~\bibnamefont {Gasparovic}}, \bibinfo
  {author} {\bibfnamefont {C.}~\bibnamefont {Broholm}}, \bibinfo {author}
  {\bibfnamefont {J.~M.~S.}\ \bibnamefont {Park}}, \bibinfo {author}
  {\bibfnamefont {A.~Y.}\ \bibnamefont {Shapiro}}, \bibinfo {author}
  {\bibfnamefont {L.~A.}\ \bibnamefont {Demianets}}, \ and\ \bibinfo {author}
  {\bibfnamefont {M.}~\bibnamefont {Kenzelmann}},\ }\href {\doibase
  10.1103/PhysRevB.88.060409} {\bibfield  {journal} {\bibinfo  {journal} {Phys.
  Rev. B}\ }\textbf {\bibinfo {volume} {88}},\ \bibinfo {pages} {060409}
  (\bibinfo {year} {2013})}\BibitemShut {NoStop}%
\bibitem [{\citenamefont {Martinoli}\ and\ \citenamefont
  {Leemann}(2000)}]{Martinoli:2000}%
  \BibitemOpen
  \bibfield  {author} {\bibinfo {author} {\bibfnamefont {P.}~\bibnamefont
  {Martinoli}}\ and\ \bibinfo {author} {\bibfnamefont {C.}~\bibnamefont
  {Leemann}},\ }\href {\doibase 10.1023/A:1004651730459} {\bibfield  {journal}
  {\bibinfo  {journal} {Journal of Low Temperature Physics}\ }\textbf {\bibinfo
  {volume} {118}},\ \bibinfo {pages} {699} (\bibinfo {year}
  {2000})}\BibitemShut {NoStop}%
\bibitem [{\citenamefont {Struck}\ \emph {et~al.}(2013)\citenamefont {Struck},
  \citenamefont {Weinberg}, \citenamefont {{\"O}lschl{\"a}ger}, \citenamefont
  {Windpassinger}, \citenamefont {Simonet}, \citenamefont {Sengstock},
  \citenamefont {H{\"o}ppner}, \citenamefont {Hauke}, \citenamefont {Eckardt},
  \citenamefont {Lewenstein},\ and\ \citenamefont {Mathey}}]{Struck:2013}%
  \BibitemOpen
  \bibfield  {author} {\bibinfo {author} {\bibfnamefont {J.}~\bibnamefont
  {Struck}}, \bibinfo {author} {\bibfnamefont {M.}~\bibnamefont {Weinberg}},
  \bibinfo {author} {\bibfnamefont {C.}~\bibnamefont {{\"O}lschl{\"a}ger}},
  \bibinfo {author} {\bibfnamefont {P.}~\bibnamefont {Windpassinger}}, \bibinfo
  {author} {\bibfnamefont {J.}~\bibnamefont {Simonet}}, \bibinfo {author}
  {\bibfnamefont {K.}~\bibnamefont {Sengstock}}, \bibinfo {author}
  {\bibfnamefont {R.}~\bibnamefont {H{\"o}ppner}}, \bibinfo {author}
  {\bibfnamefont {P.}~\bibnamefont {Hauke}}, \bibinfo {author} {\bibfnamefont
  {A.}~\bibnamefont {Eckardt}}, \bibinfo {author} {\bibfnamefont
  {M.}~\bibnamefont {Lewenstein}}, \ and\ \bibinfo {author} {\bibfnamefont
  {L.}~\bibnamefont {Mathey}},\ }\href {http://dx.doi.org/10.1038/nphys2750}
  {\bibfield  {journal} {\bibinfo  {journal} {Nature Physics}\ }\textbf
  {\bibinfo {volume} {9}},\ \bibinfo {pages} {738} (\bibinfo {year}
  {2013})}\BibitemShut {NoStop}%
\bibitem [{\citenamefont {Maryasin}\ and\ \citenamefont
  {Zhitomirsky}(2013)}]{Maryasin:2013}%
  \BibitemOpen
  \bibfield  {author} {\bibinfo {author} {\bibfnamefont {V.~S.}\ \bibnamefont
  {Maryasin}}\ and\ \bibinfo {author} {\bibfnamefont {M.~E.}\ \bibnamefont
  {Zhitomirsky}},\ }\href {\doibase 10.1103/PhysRevLett.111.247201} {\bibfield
  {journal} {\bibinfo  {journal} {Phys. Rev. Lett.}\ }\textbf {\bibinfo
  {volume} {111}},\ \bibinfo {pages} {247201} (\bibinfo {year}
  {2013})}\BibitemShut {NoStop}%
\bibitem [{\citenamefont {Smirnov}\ \emph {et~al.}(2017)\citenamefont
  {Smirnov}, \citenamefont {Soldatov}, \citenamefont {Petrenko}, \citenamefont
  {Takata}, \citenamefont {Kida}, \citenamefont {Hagiwara}, \citenamefont
  {Shapiro},\ and\ \citenamefont {Zhitomirsky}}]{Smirnov:2017}%
  \BibitemOpen
  \bibfield  {author} {\bibinfo {author} {\bibfnamefont {A.~I.}\ \bibnamefont
  {Smirnov}}, \bibinfo {author} {\bibfnamefont {T.~A.}\ \bibnamefont
  {Soldatov}}, \bibinfo {author} {\bibfnamefont {O.~A.}\ \bibnamefont
  {Petrenko}}, \bibinfo {author} {\bibfnamefont {A.}~\bibnamefont {Takata}},
  \bibinfo {author} {\bibfnamefont {T.}~\bibnamefont {Kida}}, \bibinfo {author}
  {\bibfnamefont {M.}~\bibnamefont {Hagiwara}}, \bibinfo {author}
  {\bibfnamefont {A.~Y.}\ \bibnamefont {Shapiro}}, \ and\ \bibinfo {author}
  {\bibfnamefont {M.~E.}\ \bibnamefont {Zhitomirsky}},\ }\href {\doibase
  10.1103/PhysRevLett.119.047204} {\bibfield  {journal} {\bibinfo  {journal}
  {Phys. Rev. Lett.}\ }\textbf {\bibinfo {volume} {119}},\ \bibinfo {pages}
  {047204} (\bibinfo {year} {2017})}\BibitemShut {NoStop}%
\bibitem [{\citenamefont {Goldstein}(1980)}]{Goldstein:1980}%
  \BibitemOpen
  \bibfield  {author} {\bibinfo {author} {\bibfnamefont {H.}~\bibnamefont
  {Goldstein}},\ }\href@noop {} {\emph {\bibinfo {title} {Classical
  Mechanics}}},\ \bibinfo {edition} {2nd}\ ed.\ (\bibinfo  {publisher}
  {Addison-Wesley},\ \bibinfo {year} {1980})\BibitemShut {NoStop}%
\bibitem [{\citenamefont {Carmo}(2016)}]{DoCarmo:2016}%
  \BibitemOpen
  \bibfield  {author} {\bibinfo {author} {\bibfnamefont {M.~P.~D.}\
  \bibnamefont {Carmo}},\ }\href@noop {} {\emph {\bibinfo {title} {Differential
  Geometry of Curves and Surfaces}}},\ \bibinfo {edition} {2nd}\ ed.\ (\bibinfo
   {publisher} {Dover},\ \bibinfo {year} {2016})\BibitemShut {NoStop}%
\bibitem [{\citenamefont {Whitham}(1974)}]{Whitham:1974}%
  \BibitemOpen
  \bibfield  {author} {\bibinfo {author} {\bibfnamefont {G.}~\bibnamefont
  {Whitham}},\ }\href@noop {} {\emph {\bibinfo {title} {Linear and nonlinear
  waves}}}\ (\bibinfo  {publisher} {Wiley-Interscience},\ \bibinfo {year}
  {1974})\BibitemShut {NoStop}%
\bibitem [{\citenamefont {Eriksson}\ \emph {et~al.}(2017)\citenamefont
  {Eriksson}, \citenamefont {Bergman}, \citenamefont {Bergqvist},\ and\
  \citenamefont {Hellsvik}}]{Eriksson:2017}%
  \BibitemOpen
  \bibfield  {author} {\bibinfo {author} {\bibfnamefont {O.}~\bibnamefont
  {Eriksson}}, \bibinfo {author} {\bibfnamefont {A.}~\bibnamefont {Bergman}},
  \bibinfo {author} {\bibfnamefont {L.}~\bibnamefont {Bergqvist}}, \ and\
  \bibinfo {author} {\bibfnamefont {J.}~\bibnamefont {Hellsvik}},\ }\href@noop
  {} {\emph {\bibinfo {title} {Atomistic Spin Dynamics: Foundations and
  Applications}}}\ (\bibinfo  {publisher} {Oxford},\ \bibinfo {year}
  {2017})\BibitemShut {NoStop}%
\bibitem [{\citenamefont {Loft}\ and\ \citenamefont
  {DeGrand}(1987)}]{Loft:1987}%
  \BibitemOpen
  \bibfield  {author} {\bibinfo {author} {\bibfnamefont {R.}~\bibnamefont
  {Loft}}\ and\ \bibinfo {author} {\bibfnamefont {T.~A.}\ \bibnamefont
  {DeGrand}},\ }\href {\doibase 10.1103/PhysRevB.35.8528} {\bibfield  {journal}
  {\bibinfo  {journal} {Phys. Rev. B}\ }\textbf {\bibinfo {volume} {35}},\
  \bibinfo {pages} {8528} (\bibinfo {year} {1987})}\BibitemShut {NoStop}%
\bibitem [{\citenamefont {Bussi}\ and\ \citenamefont
  {Parrinello}(2007)}]{Bussi:2007}%
  \BibitemOpen
  \bibfield  {author} {\bibinfo {author} {\bibfnamefont {G.}~\bibnamefont
  {Bussi}}\ and\ \bibinfo {author} {\bibfnamefont {M.}~\bibnamefont
  {Parrinello}},\ }\href {\doibase 10.1103/PhysRevE.75.056707} {\bibfield
  {journal} {\bibinfo  {journal} {Phys. Rev. E}\ }\textbf {\bibinfo {volume}
  {75}},\ \bibinfo {pages} {056707} (\bibinfo {year} {2007})}\BibitemShut
  {NoStop}%
\bibitem [{Sup()}]{SupplMat}%
  \BibitemOpen
  \href@noop {} {}\bibinfo {howpublished} {http://xxxx.xxx}\BibitemShut
  {NoStop}%
\bibitem [{\citenamefont {Ralko}\ \emph {et~al.}(2008)\citenamefont {Ralko},
  \citenamefont {Poilblanc},\ and\ \citenamefont {Moessner}}]{Ralko:2008}%
  \BibitemOpen
  \bibfield  {author} {\bibinfo {author} {\bibfnamefont {A.}~\bibnamefont
  {Ralko}}, \bibinfo {author} {\bibfnamefont {D.}~\bibnamefont {Poilblanc}}, \
  and\ \bibinfo {author} {\bibfnamefont {R.}~\bibnamefont {Moessner}},\ }\href
  {\doibase 10.1103/PhysRevLett.100.037201} {\bibfield  {journal} {\bibinfo
  {journal} {Phys. Rev. Lett.}\ }\textbf {\bibinfo {volume} {100}},\ \bibinfo
  {pages} {037201} (\bibinfo {year} {2008})}\BibitemShut {NoStop}%
\bibitem [{\citenamefont {Hayashi}\ \emph {et~al.}(2014)\citenamefont
  {Hayashi}, \citenamefont {Nawata}, \citenamefont {Taira}, \citenamefont
  {Shikata}, \citenamefont {Kawase},\ and\ \citenamefont
  {Minamide}}]{Hayashi:2014}%
  \BibitemOpen
  \bibfield  {author} {\bibinfo {author} {\bibfnamefont {S.}~\bibnamefont
  {Hayashi}}, \bibinfo {author} {\bibfnamefont {K.}~\bibnamefont {Nawata}},
  \bibinfo {author} {\bibfnamefont {T.}~\bibnamefont {Taira}}, \bibinfo
  {author} {\bibfnamefont {J.-i.}\ \bibnamefont {Shikata}}, \bibinfo {author}
  {\bibfnamefont {K.}~\bibnamefont {Kawase}}, \ and\ \bibinfo {author}
  {\bibfnamefont {H.}~\bibnamefont {Minamide}},\ }\href
  {http://dx.doi.org/10.1038/srep05045} {\bibfield  {journal} {\bibinfo
  {journal} {Scientific Reports}\ }\textbf {\bibinfo {volume} {4}},\ \bibinfo
  {pages} {5045 EP } (\bibinfo {year} {2014})}\BibitemShut {NoStop}%
\bibitem [{\citenamefont {Kawamura}\ and\ \citenamefont
  {Miyashita}(1985)}]{Kawamura:1985}%
  \BibitemOpen
  \bibfield  {author} {\bibinfo {author} {\bibfnamefont {H.}~\bibnamefont
  {Kawamura}}\ and\ \bibinfo {author} {\bibfnamefont {S.}~\bibnamefont
  {Miyashita}},\ }\href {\doibase 10.1143/JPSJ.54.4530} {\bibfield  {journal}
  {\bibinfo  {journal} {Journal of the Physical Society of Japan}\ }\textbf
  {\bibinfo {volume} {54}},\ \bibinfo {pages} {4530} (\bibinfo {year}
  {1985})}\BibitemShut {NoStop}%
\bibitem [{\citenamefont {Gvozdikova}\ \emph {et~al.}(2011)\citenamefont
  {Gvozdikova}, \citenamefont {Melchy},\ and\ \citenamefont
  {Zhitomirsky}}]{Gvozdikova:2011}%
  \BibitemOpen
  \bibfield  {author} {\bibinfo {author} {\bibfnamefont {M.~V.}\ \bibnamefont
  {Gvozdikova}}, \bibinfo {author} {\bibfnamefont {P.-E.}\ \bibnamefont
  {Melchy}}, \ and\ \bibinfo {author} {\bibfnamefont {M.~E.}\ \bibnamefont
  {Zhitomirsky}},\ }\href {http://stacks.iop.org/0953-8984/23/i=16/a=164209}
  {\bibfield  {journal} {\bibinfo  {journal} {Journal of Physics: Condensed
  Matter}\ }\textbf {\bibinfo {volume} {23}},\ \bibinfo {pages} {164209}
  (\bibinfo {year} {2011})}\BibitemShut {NoStop}%
\bibitem [{\citenamefont {Seabra}\ \emph {et~al.}(2011)\citenamefont {Seabra},
  \citenamefont {Momoi}, \citenamefont {Sindzingre},\ and\ \citenamefont
  {Shannon}}]{Seabra:2011}%
  \BibitemOpen
  \bibfield  {author} {\bibinfo {author} {\bibfnamefont {L.}~\bibnamefont
  {Seabra}}, \bibinfo {author} {\bibfnamefont {T.}~\bibnamefont {Momoi}},
  \bibinfo {author} {\bibfnamefont {P.}~\bibnamefont {Sindzingre}}, \ and\
  \bibinfo {author} {\bibfnamefont {N.}~\bibnamefont {Shannon}},\ }\href
  {\doibase 10.1103/PhysRevB.84.214418} {\bibfield  {journal} {\bibinfo
  {journal} {Phys. Rev. B}\ }\textbf {\bibinfo {volume} {84}},\ \bibinfo
  {pages} {214418} (\bibinfo {year} {2011})}\BibitemShut {NoStop}%
\end{thebibliography}%

\end{document}